\documentclass[a4paper,11pt]{article}
\usepackage{jcappub}
\usepackage[utf8]{inputenc}
\usepackage{amsmath,mathtools}

\usepackage{tensor}
\usepackage{amssymb}
\usepackage{graphicx}
\usepackage{tikz}
\usepackage{physics}
\usepackage{lipsum}  
\usepackage{bigints}
\usepackage{verbatim}
\usepackage{bbold}
\usepackage[normalem]{ulem}
\newcommand{\leri}[1]{\left(#1\right)}

\usepackage{subcaption}

\title{Landau damping for gravitational waves in parity-violating theories}

\author[a,1]{F. Bombacigno,\note{Corresponding author.}}
\author[a]{F. Moretti,}
\author[b,c]{S. Boudet,}
\author[a,d]{Gonzalo J. Olmo,}

\affiliation[a]{Departament de F\'{i}sica Teòrica and IFIC, Centro Mixto Universitat de València - CSIC, Universitat de València, Burjassot 46100, València, Spain}
\affiliation[b]{Dipartimento di Fisica, Universit\`{a} di Trento,\\Via Sommarive 14, I-38123 Povo (TN), Italy}
\affiliation[c]{Trento Institute for Fundamental Physics and Applications (TIFPA)-INFN,\\Via Sommarive 14, I-38123 Povo (TN), Italy}
\affiliation[d]{Universidade Federal do Cear\'a (UFC), Departamento de F\'isica,\\ Campus do Pici, Fortaleza - CE, C.P. 6030, 60455-760 - Brazil.}

\emailAdd{flavio2.bombacigno@uv.es}
\emailAdd{fabio.moretti@ext.uv.es}
\emailAdd{simon.boudet@unitn.it}
\emailAdd{gonzalo.olmo@uv.es}

\abstract{We discuss how tensor polarizations of gravitational waves can suffer Landau damping in the presence of velocity birefringence, when parity symmetry is explicitly broken. In particular, we analyze the role of the Nieh-Yan and Chern-Simons terms in modified theories of gravity, showing how the gravitational perturbation in collisionless media can be characterized by a subluminal phase velocity, circumventing the well-known results of General Relativity and allowing for the appearance of the kinematic damping. We investigate in detail the connection between the thermodynamic properties of the medium, such as temperature and mass of the particles interacting with the gravitational wave, and the parameters ruling the parity violating terms of the models.
In this respect, we outline how the dispersion relations can give rise in each model to different regions of the wavenumber space, where the phase velocity is subluminal, superluminal or does not exist. Quantitative estimates on the considered models indicate that the phenomenon of Landau damping is not detectable given the sensitivity of present-day instruments.}

\makeatletter
\gdef\@fpheader{}
\makeatother
\begin{document}

\maketitle
\flushbottom
\section{Introduction}\label{intro}
In the last years many alternative theories of gravity have been formulated with the aim of providing consistent explanations to astrophysical and cosmological phenomena for which General Relativity (GR) yields unsatisfactory predictions, like for instance the accelerated expansion of the Universe and the problem of the dark matter \cite{DeFelice:2010aj,Nojiri:2010wj,Olmo:2011uz,Cai:2015emx,NOJIRI20171,Krssak:2018ywd,Olmo:2019flu,Cabral:2020fax,Harko:2020ibn,Capozziello:2022lic,Fernandes:2022zrq}, or more recently the so called $H_0$ tension \cite{Yan:2019gbw,SolaPeracaula:2020vpg,Schoneberg:2021qvd,DiValentino:2021izs,Dainotti:2021pqg}. It is thereby fundamental to derive phenomenological signatures that enable us to compare observations with the predictions obtained from relevant alternative models. In the current landscape of observational methods and techniques, gravitational wave astronomy stands as one of the most valuable and promising avenues through which to perform tests on gravitational theories. In particular, the forthcoming introduction of 3G detectors, such as Cosmic Explorer and Einstein Telescope, will allow us to detect binary black hole mergers up to a redshift $z\sim 30$ \cite{Ng:2020qpk}, covering the whole length of the stellar era and greatly improving our ability to resolve between distinct polarizations \cite{Takeda:2019gwk,Isi:2022mbx}. It is known, indeed, that the tensor, vector, and scalar modes carried by the gravitational degrees of freedom interact in different ways with the sources and the traversed media. We refer for instance to the attenuation of the gravitational signal in the presence of a dissipative fluid \cite{Hawking:1966qi,Madore,Madore:1972ww,Prasanna:1999pn,Anile}, an expanding Universe \cite{1978SvA....22..528Z,Weinberg:2003ur,Flauger:2017ged} and a cosmological neutrino background \cite{Lattanzi:2005xb,Lattanzi:2010gn,Benini:2010zz}, or to the role of non-tensor polarizations in the rate of energy loss for binary orbital motions where modified theories of gravity are considered \cite{Krause:1994ar,Zhang:2017srh,Brito:2015oca,LIGOScientific:2018dkp,Wagle:2018tyk}. 

In this work, in particular, we deal with the so called gravitational Landau damping, consisting in the kinematic damping, or enhancement, of the metric perturbation during its propagation in a non-collisional medium. A number of works in the literature have analyzed the possibility of emergence of such phenomenon in the case of gravitational waves in GR, considering either a flat Minkowski or an expanding FLRW background  \cite{Chesters:1973wan,PhysRevD.13.2724,Gayer:1979ff,Weinberg:2003ur,Lattanzi:2005xb,Benini:2010zz,Flauger:2017ged,Baym:2017xvh}. With great generality one can affirm that Landau damping is possible for tensor gravitational waves in GR only by including anisotropies in the medium background configuration (perturbations on Minkowski spacetime), or by considering the coupling between wave perturbations and background curvature (FLRW case). On the contrary, in the simplest case of gravitational waves in GR traveling on a static flat spacetime and interacting with a collisionless medium with isotropic background configuration, Landau damping is forbidden. Indeed, as shown by the explicit calculation of the dispersion relation, gravitational modes within the medium are characterized by superluminal phase velocity, which can be demonstrated to be the sufficient condition for the non-existence of this phenomenon.

To broaden our understanding on  Landau damping phenomenology, here we are instead interested in studying the interaction between tensor gravitational waves in the context of alternative theories of gravity, and non-collisional matter with isotropic background configuration on a Minkowski background. This latter choice is motivated by the fact that setting our analysis on a flat metric background enables us to investigate the intrinsic properties of the modified gravity model considered. Moreover, as shown in \cite{Garg:2021baw,Garg:2022wdm}, taking into account the background curvature generated by the medium itself, which can be in principle calculated by solving the Einstein equations for the unperturbed matter distribution $f_0$, would return a wrong ordering of perturbations, i.e. the inclusion of extra-terms in the self-consistent equations governing the wave dynamics within the medium (we refer to equations \eqref{sceqny}, \eqref{sceqcs} and \eqref{sceqcsaff}) that should instead be discarded in the context of a linearized analysis.\footnote{As can be noticed from \eqref{T left def} and \eqref{T right def}, the matter polarization tensors we obtain with explicit calculations are already expressed at first order in the metric perturbation.}
In addition to this, the study we perform is grounded on the hypothesis of homogeneous and isotropic equilibrium configuration of the particles, which is satisfied only by considering a small length scale of gravitational radiation with respect to the total size of the medium. For waves characterized by such wavelengths, the medium appears indeed as infinite, and the background metric can always be made quasi-Minkowskian (up to irrelevant Newtonian and post-Newtonian corrections) by a suitable choice of a freely falling frame on a sufficiently large patch of space-time on which our analysis will concentrate.

In \cite{Moretti:2020kpp,Moretti:2021ljj} we demonstrated that the occurrence of the Landau damping is ultimately driven by the structure of the linearized equations of motion, which determines the nature of the wave operator acting on the different polarization states, and the relevant stress energy tensor components appearing at the source. In particular, we showed how in Horndeski theories of gravity, where the propagation of the additional scalar mode is encoded in a massive Klein-Gordon equation sourced by the trace of the stress energy tensor, Landau damping occurs if the mass of the scalar mode and the thermodynamic properties of the medium satisfy a typical relation (see inequality (33) of \cite{Moretti:2020kpp}). In this case, indeed, propagation is characterized by a subluminal phase velocity for all the wavenumbers $k$. Always in \cite{Moretti:2020kpp}, we also showed that tensor modes, which in Horndeski theories are still described on a Minkowski background\footnote{For cosmological effects see for example \cite{Lombriser:2015sxa,Kobayashi:2019hrl,Crisostomi:2016czh}.} by a GR-like equation (up to a redefinition of the Newton constant), cannot suffer kinematic damping, being their propagation in the medium characterized instead by a superluminal phase velocity, as it occurs in GR.

It is clear, then, that in order tensor polarizations experience Landau damping, we need an equation of motion able to give rise in matter to a deformed dispersion relation with respect to the GR case. This can be achieved either by operating on the source terms, as in the non-minimal curvature-matter coupling $f(R,\mathcal{L}_m)$ theories\footnote{That will be the subject of a forthcoming work \cite{bombacigno2023}.} \cite{Harko:2010mv,Harko:2011kv,Barrientos:2018cnx}, or by modifying the differential structure of the equation of motion, looking for theories where the standard d'Alembert operator is supplemented by additional terms. In this respect, a possible choice is represented by theories of gravity exhibiting parity violation, where the propagation of the two circular polarization modes (left and right handed states \cite{Isi:2022mbx}), is described by different equations of motion. This leads to the so called gravitational birefringence \cite{Conroy:2019ibo,Qiao:2019wsh,Zhao:2019xmm,Chatzistavrakidis:2021oyp,Hohmann:2022wrk,Wu:2021ndf,Martin-Ruiz:2017cjt,Nojiri:2019nar,Nojiri:2020pqr,Boudet:2022nub,Li:2022grj}, which consists in different behaviour for the amplitude and the phase/group velocity of each polarization state. In particular, phase velocity is usually affected in such a way that its expression for left and right modes differs from GR for corrections of opposite sign. This naturally generates a branch of solutions with subluminal phase velocity, which as previously discussed, represents the condition for Landau damping to be in principle possible. Theories characterized by birefringence emerge, therefore, as a candidate where to investigate the possibility that tensor modes could interact non-collisionally with matter.
Parity-violating theories are recently receiving increasing attention for their role in modern physics topics such as CMB polarization \cite{ALEXANDER2008444,PhysRevLett.83.1506,Bartolo:2018elp,Bartolo:2017szm}, gravitational waves \cite{PhysRevD.99.064049,Odintsov:2021kup,sym14040729,PhysRevD.105.104054,Satoh:2007gn,Takahashi:2009wc,Kamada:2021kxi}, baryon asymmetry problem \cite{PhysRevLett.96.081301,PhysRevD.69.023504,Alexander_2006,Jimenez:2017cdr} and black hole perturbations \cite{PhysRevD.80.064008,PhysRevD.81.089903,PhysRevD.81.124021,PhysRevD.80.064006,Yunes:2010yf,Cano:2019ore,Harko:2009kj,Ghodrati:2017roz,Motohashi:2011pw,Yagi:2017zhb,Wagle:2021tam}. Among the possible formulations, we mention for instance Holst and Nieh-Yan extensions \cite{Iosifidis:2020dck,Li:2022vtn}, degenerate higher-order scalar-tensor theories (DHOST) \cite{Qiao:2019wsh}, Chern-Simons modified gravity (CSMG) \cite{Jackiw:2003pm,Alexander:2009tp,Boudet:2022wmb,Sulantay:2022sag}, bumblebee models \cite{Kostelecky:2003fs,Bluhm:2004ep,Bailey:2006fd,Bluhm:2007bd,Delhom:2019wcm,Delhom:2020gfv,Delhom:2022xfo} and Hořava-Lifshitz gravity \cite{Zhu:2013fja,Gong:2021jgg}. 

Motivated by the phenomenon of birefringence, in this work we ascertain if two specific classes of parity violating theories can be actually affected by Landau damping. We focus, specifically, on Nieh-Yan (NY) models as they are formulated in teleparallel gravity \cite{Qiao:2019wsh,Zhao:2019xmm,Conroy:2019ibo,Li:2020xjt,Li:2021wij,Wu:2021ndf,Chatzistavrakidis:2021oyp}, and CSMG, that we analyze both from the metric perspective \cite{Jackiw:2003pm,Alexander:2009tp} and adopting a metric-affine approach, recently developed in \cite{Boudet:2022wmb,Boudet:2022nub}. This choice enables us to analyze the effects of parity violating terms in two different scenarios, where birefringence is already present in the vacuum (NY) or only appears when the coupling with matter is considered (CSMG). This is due to the nature of the corrections introduced in the equation for the metric perturbation when a Minkowskian background is considered. In the NY theory, indeed, the d'Alembert operator comes with an additional term carrying first-order derivatives, whose sign depends on the polarization state. This implies that the dispersion relations for left and right modes are distinguished even in vacuum, leading to birefringence. Conversely, in CSMG corrections to the wave operator can be collected in a global multiplicative factor containing first spatial derivatives, which appear with opposite sign according to the polarization considered. This indicates that the general solution in vacuum is the superposition of a standard propagating wave and a spatial oscillation of fixed wavelength, which is determined by the parameters quantifying the parity violation. Such an oscillation is however related with the non-radiating component of the metric perturbation, so that in vacuum we eventually retain only the wave described by the dispersion relation of the GR. Analogous results hold also when the CSMG is addressed from a metric-affine perspective, consisting in considering the metric and the affine connection as a priori independent quantities. In this case, indeed, metric perturbations are accompanied with torsion perturbations, and the analysis in the Fourier-Laplace space points out that birefringence appears again only in the presence of matter.

A common prediction to all the scenarios is the existence of specific regions in the wavenumber space where Landau damping is possible, together with intervals where free propagation within the medium is forbidden. Their extension is in general determined by the thermodynamic properties of the particles interacting with the gravitational perturbation, like density and temperature, as well as by the parameters ruling the parity breaking terms, which concur in defining the peculiar threshold values of the wavenumber $k$ separating the domains where the phase velocity of the wave is subluminal, superluminal, or not existing.
 
The paper is organized as follows. In sec.~\ref{sec2} we analyze the propagation of gravitational waves in teleparallel NY theories and we introduce the main formalism required for dealing with the Landau damping phenomenon. In sec.~\ref{sec3} and sec.~\ref{sec4} we repeat the analysis for CSMG, turning our attention to the metric and metric-affine formulation, respectively,  outlining the analogies and the differences between the two approaches. In sec.~\ref{sec5} we estimate, for the three cases, the amount of damping suffered by a typical signal traveling within a dark matter halo in an environment compatible with Solar System conditions. Eventually, conclusions are drawn in sec.~\ref{sec6}.

The metric perturbation $h_{\mu\nu}$ on the Minkowski background is defined in the local chart $g_{\mu\nu}=\eta_{\mu\nu}+h_{\mu\nu}$, with $h_{\mu\nu}\ll 1$, where for the Minkowski metric $\eta_{\mu\nu}$ the mostly plus signature is chosen. Gravitational coupling is set as $\chi=8\pi$, using geometrized units $G=c=1$. Boltzmann constant is set to unity, i.e. $k_B=1$. The Levi-Civita tensor $\varepsilon_{\mu\nu\rho\sigma}$  is defined in terms of the completely antisymmetric symbol $\epsilon_{\mu\nu\rho\sigma}$, with $\epsilon_{0123}=1$.

\section{Teleparallel Nieh-Yan gravity}\label{sec2}
As discussed in \cite{Qiao:2019wsh,Zhao:2019xmm,Li:2020xjt,Li:2021wij,Wu:2021ndf,Chatzistavrakidis:2021oyp} in the context of the teleparallel formulation (see also the symmetric teleparallel case \cite{Conroy:2019ibo}), Nieh-Yan parity-violating models lead to the following modified equation for the metric perturbation on a Minkowskian background: 
\begin{equation}
    \Box h_{\mu\nu}-\alpha \epsilon_{\rho\sigma\lambda(\mu}\partial^\rho\theta\partial^\sigma h\indices{^\lambda_{\nu)}}=-2\chi T_{\mu\nu},
\label{NY equation metric}
\end{equation}
where $\alpha$ is a coupling constant keeping trace of the additional parity-violating terms in the action, and $T_{\mu\nu}$ is the stress energy tensor perturbation. According to the standard approach, the pseudoscalar field $\theta$ is assumed to depend, adiabatically, only on the cosmological time, so that it can be considered nearly constant during the propagation of the gravitational perturbations. This amounts to set the time derivative of the $\theta$ field constant, which we define as $\dot{\theta}\simeq \dot{\theta}_B$, where the subscript $B$ stands for background. Accordingly, second order derivatives are instead neglected. The evolution of tensor modes for a gravitational wave travelling along the $z$ axis is then displayed by
\begin{align}
    &\Box h_+-\frac{\alpha\dot{\theta}_B}{2} \partial_z h_\times=-2\chi T_{11},
    \label{equation tensor modes 1}
    \\
    &\Box h_\times+\frac{\alpha\dot{\theta}_B}{2} \partial_z h_+=-2\chi T_{12},
\label{equation tensor modes 2}
\end{align}
where the metric perturbation is set as
\begin{equation}\label{metric perturbation tt gauge}
 h_{ij} = \begin{pmatrix}
 h_+ & h_\times & 0 \\
 h_\times & -h_+ & 0 \\
 0 & 0 & 0
\end{pmatrix}.
\end{equation}
Plus and cross polarizations can be then decoupled by introducing the circularized right and left modes \cite{Isi:2022mbx}, defined by:
\begin{equation}
    \begin{cases}
    &h_R=\frac{h_++i h_\times}{\sqrt{2}}\\
    &h_L=\frac{h_+-i h_\times}{\sqrt{2}}
    \end{cases}
    \iff
    \begin{cases}
    &h_+=\frac{h_R+h_L}{\sqrt{2}}\\
    &h_\times=\frac{h_R-h_L}{i\sqrt{2}},
    \end{cases}
\end{equation}
which allow us to recast \eqref{equation tensor modes 1}-\eqref{equation tensor modes 2} as
\begin{align}
    &\leri{\Box-\frac{i\alpha\dot{\theta}_B}{2}\partial_z}h_L=-2\chi\frac{T_{11}-i T_{12}}{\sqrt{2}}\equiv -2\chi T_L,
\label{equation tensor chiral modes TEGR L}
    \\
    &\leri{\Box+\frac{i\alpha\dot{\theta}_B}{2}\partial_z}h_R=-2\chi\frac{T_{11}+i T_{12}}{\sqrt{2}}\equiv -2\chi T_R.
\label{equation tensor chiral modes TEGR R}
\end{align}
As outlined in \cite{Chatzistavrakidis:2021oyp}, left and right modes are characterized in vacuum by the dispersion relation:
\begin{equation}
    \mathcal{D}_{L,R}(\omega,k)=\omega^2-k^2\pm \frac{\alpha\dot{\theta}_B}{2}k=0,
\end{equation}
so that group and phase velocities result in
\begin{equation}
    v_g\equiv\frac{d\omega}{d k}=\frac{1\mp\frac{\alpha\dot{\theta}_B }{4k}}{\sqrt{1\mp\frac{\alpha\dot{\theta}_B }{2k}}},\qquad v_p\equiv\frac{\omega}{k}=\sqrt{1\mp\frac{\alpha\dot{\theta}_B }{2k}}.
\end{equation}
\\For what concerns the propagation in matter we begin our analysis by considering a relativistic medium composed by neutral particles of the same mass $m$, described in the context of the kinetic theory by a probability distribution function $f(\vec{x},\vec{p},t)$. Specifically, we consider a function defined in the single-particle phase space\footnote{Following \cite{Weinberg:2003ur,Flauger:2017ged} we take as canonical coordinates in the single particle phase space the contravariant components of the position vector $x^i$ and the covariant components of the momentum $p_i$.} and properly normalized in order to return the total number of particles $N$ when integrated over its entire domain. This implies that the number $dN$ of particles with positions between $\vec{x}$ and $\vec{x}+d\vec{x}$ and momenta between $\vec{p}$ and $\vec{p}+d\vec{p}$ is given by $dN=f(\vec{x},\vec{p},t)d\vec{x}\, d\vec{p}$. Before the passage of the gravitational perturbation, we assume that the medium has reached an equilibrium state, described by some distribution $f_0$, characterized by a temperature $T$. We remark again that our analysis is focused on gravitational radiation with wavelength much smaller than the length scale of variation of the thermodynamic properties of the medium, like density, pressure and temperature. Under this assumption the background configuration of the particles can be reasonably well described by a homogeneous and isotropic distribution function. Then, in order to include relativistic effects for the massive particles of the medium, we fix the equilibrium configuration as a J\"{u}ttner-Maxwell distribution
\begin{equation}
    f_0(p)=\dfrac{n}{4 \pi m^2 T K_2 \leri{x}}e^{-\frac{\sqrt{m^2+p^2}}{T}},
    \label{max jutt}
\end{equation}
where we introduced the density of particles $n$, the modulus of the flat three momentum $p=\sqrt{\delta^{ij}p_i p_j}$, and the modified Bessel function of the second kind $K_\nu\leri{\cdot}$ of real index $\nu$, evaluated in $x\equiv \frac{m}{T}$.
The evolution of the medium is provided by the Vlasov equation for the distribution function $f(\vec{x},\vec{p},t)$, i.e.
\begin{equation}
  \dfrac{D f}{dt}=\dfrac{\partial f}{\partial t}+\dfrac{d x^m}{dt}\dfrac{\partial f}{\partial x^m}+\dfrac{dp_m}{dt}\dfrac{\partial f}{\partial p_m}=0,
\end{equation}
the latter being nothing more than Boltzmann equation in which the collision integral at right-handed side is neglected. It must be stressed that the assumption of a collisionless medium can be made when the mean free path of the particles is much greater than the total size of the system or, alternatively, when the rate of collisions is much smaller than $\frac{1}{\Delta T}$, being $\Delta T$ the global time-scale of observation, i.e. the total time of interaction between the gravitational waves and the medium in the scenario here considered.
Then, by means of the identity $\frac{d x^i}{dt}=\frac{p^i}{p^0}$, with $p^0=\sqrt{m^2+g^{ij}p_ip_j}$ representing the energy of the particle, the geodesic equation can be rewritten as
\begin{equation}
    \frac{dp^m}{dt}+\Gamma\indices{^m_{\alpha\beta}}\frac{p^\alpha p^\beta}{p^0}=0.
\end{equation}
It follows that the time derivative of $p_m$ is given by
\begin{equation}
    \dfrac{dp_m}{dt}=\frac{\partial g_{mn}}{\partial t}p^n+\frac{\partial g_{mn}}{\partial x^k}\frac{p^k p^n}{p^0}-g_{mn}\Gamma\indices{^n_{\alpha\beta}}\frac{p^\alpha p^\beta}{p^0},
\end{equation}
which taking into account the perturbative expansion $g_{\mu\nu}=\eta_{\mu\nu}+h_{\mu\nu}$ reduces to
\begin{equation}
    \dfrac{dp_m}{dt}=\dfrac{p^ip^j}{2p^0}\dfrac{\partial g_{ij}}{\partial x^m}=\dfrac{p_ip_j}{2p^0}\dfrac{\partial h_{ij}}{\partial x^m},
\end{equation}
where we simply lowered the indices of $p^i$ with $\eta_{ij}$, as $\frac{dp_m}{dt}$ is already of first order in perturbation. 
\\ \noindent Before the arrival of the gravitational wave, the distribution function is assumed to be some isotropic equilibrium solution $f_0\leri{p}$ of the unperturbed equation. Therefore, at the initial time $t=0$ we simply have $f(\vec{x},\vec{p},0)=f_0( \sqrt{g^{ij}(\vec{x},0)p_ip_j})$, which at the first order in perturbation results in 
\begin{equation}
   f(\vec{x},\vec{p},0)=f_0 \leri{p}-\dfrac{f_0'(p)}{2}\dfrac{p_ip_j}{p}h_{ij}(\vec{x},0),
\end{equation}
where $f_0'(p) \equiv \frac{d f_0}{dp}$ and we used the fact that at the first order $p_ip_j h^{ij}=p_ip_j h_{ij}$.
\\For $t>0$, the distribution function is perturbed by the gravitational wave, i.e.
\begin{equation}
   f(\vec{x},\vec{p},t)=f_0 \leri{p}-\dfrac{f_0'(p)}{2}\dfrac{p_ip_j}{p}h_{ij}(\vec{x},t)+\delta f (\vec{x},\vec{p},t),
\end{equation}
where $\delta f (\vec{x},\vec{p},t)$ is small with respect to the equilibrium configuration, that is $\frac{\delta f}{f_0}=\mathcal{O}(h)$. The linearized Vlasov equation for the perturbation $\delta f (\vec{x},\vec{p},t)$ is then given by:
\begin{equation}
    \frac{\partial \delta f }{\partial t}+\frac{p^m}{p^0}\frac{\partial \delta f}{\partial x^m}-\frac{f_0'(p) }{2p} \frac{\partial h_{ij}}{\partial t}p_ip_j=0.
  \label{linear vlasov}
\end{equation}
In order for the dynamical problem to be well posed, this last equation has to be accompanied by the equations for the metric perturbation \eqref{equation tensor chiral modes TEGR L}-\eqref{equation tensor chiral modes TEGR R}, with the sources conveniently rewritten in terms of the distribution function $f(\vec{x},\vec{p},t)$. That can be done by means of the stress energy tensor of a Vlasov gas, i.e.
\begin{equation}
    T_{ij}(\vec{x},t)=\frac{1}{\sqrt{-g}}\int d^3p\, \dfrac{p_ip_j}{p^0}f (\vec{x},\vec{p},t),
\end{equation}
which at first order on a Minkowski background is simply given by
\begin{equation}
T_{ij}(\vec{x},t)=\int d^3p\, \dfrac{p_ip_j}{p^0}\delta f (\vec{x},\vec{p},t),
\end{equation}
where\footnote{In the following we will always use the notation $d^3p$ instead of $d\vec{p}$, previously introduced.} $d^3p=dp_1dp_2dp_3$. The set of equations formed by \eqref{equation tensor chiral modes TEGR L}, \eqref{equation tensor chiral modes TEGR R} and \eqref{linear vlasov} represents a system of coupled differential equations for $h_{L,R}(t,z)$ and $f(\vec{x},\vec{p},t)$. However, by performing a Fourier transform on spatial coordinates and a Laplace transform on time $t$, it can be simply converted into an algebraic problem. The perturbation $\delta f(\vec{x},\vec{p},t)$ is therefore obtained by solving \eqref{linear vlasov} and it reads
\begin{equation}
    \delta f^{(k,s)}(\vec{p})=\frac{\frac{f_0'(p)}{2 p}\leri{s\,h_{ij}^{(k,s)}-h_{ij}^{(k)}(0)}p_ip_j}{s+ik \frac{p_3}{p^0}},
\label{deltafFL}
\end{equation}
where the Fourier and Fourier-Laplace components of a generic field are displayed as $\phi^{(k)}(t)$ and $\phi^{(k,s)}$, respectively. It follows that the sources for the right and left modes can be expressed as
\begin{align}
    T_L= \frac{1}{4}&\int d^3p \, \frac{f_0'(p)p_1^2}{p\leri{p^0s+ikp_3}}\left [ \leri{p_1^2-3p_2^2}\leri{s h_R^{(k,s)}-h_R^k(0)}+
    \leri{p_1^2+p_2^2}\leri{s h_L^{(k,s)}-h_L^k(0)}\right ]
    \label{T left}
    \\
    T_R= \frac{1}{4}&\int d^3p \, \frac{f_0'(p)p_1^2}{p\leri{p^0s+ikp_3}}\left [ \leri{p_1^2-3p_2^2}\leri{s h_L^{(k,s)}-h_L^k(0)}+
    \leri{p_1^2+p_2^2}\leri{s h_R^{(k,s)}-h_R^k(0)}\right ],
    \label{T right}
\end{align}
where we neglected terms odd in $p_1,p_2$ which are identically vanishing on symmetry grounds. Now, by introducing cylindrical coordinates in the momentum space, i.e.
\begin{equation}
    p_1=\rho \cos \varphi,\quad p_2=\rho \sin \varphi,\quad p_3=p_3,
\end{equation}
the sources \eqref{T left}-\eqref{T right} can be rewritten as
\begin{align}
    T_L= \frac{1}{4}&\int dV \, \frac{f_0'(p)\rho^5\cos^2\varphi}{p\leri{p^0s+ikp_3}}\left [ \leri{4\cos^2\varphi-3}\leri{s h_R^{(k,s)}-h_R^k(0)}+\leri{s h_L^{(k,s)}-h_L^k(0)}\right ]
    \label{T left cyl}
    \\
    T_R= \frac{1}{4}&\int dV \, \frac{f_0'(p)\rho^5\cos^2\varphi}{p\leri{p^0s+ikp_3}}\left [ \leri{4\cos^2\varphi-3}\leri{s h_L^{(k,s)}-h_L^k(0)}+\leri{s h_R^{(k,s)}-h_R^k(0)}\right ],
    \label{T right cyl}
\end{align}
where now integration is performed in the volume $dV=d\varphi d\rho dp_3$, with:
\begin{equation}
    \varphi\in[0,2\pi],\;\rho\in[0,+\infty),\;p_3\in(-\infty,+\infty).
\end{equation}
The angular parts of the integrals corresponding to the different polarization states can be evaluated separately, and simply result in
\begin{equation}
    \int_0^{2\pi}d\varphi \;(4\cos^2\varphi-3)\cos^2\varphi = 0,\quad  \int_0^{2\pi}d\varphi \cos^2\varphi = \pi,
\end{equation}
which implies that left and right modes are not mixed by the sources, i.e.
\begin{align}
    T_L= \frac{\pi}{4}&\int_0^\infty d\rho \int_{-\infty}^{+\infty}dp_3 \, \frac{f_0'(p)\rho^5}{p\leri{p^0s+ikp_3}}\leri{s h_L^{(k,s)}-h_L^k(0)}
    \label{T left def}
    \\
    T_R= \frac{\pi}{4}&\int_0^\infty d\rho \int_{-\infty}^{+\infty}dp_3 \, \frac{f_0'(p)\rho^5}{p\leri{p^0s+ikp_3}}\leri{s h_R^{(k,s)}-h_R^k(0)}.
    \label{T right def}
\end{align}
We can then explicitly solve \eqref{equation tensor chiral modes TEGR L}-\eqref{equation tensor chiral modes TEGR R} for the Fourier-Laplace components of the gravitational wave, i.e.
\begin{equation}\label{sceqny}
    h_{L,R}^{(k,s)}=\dfrac{\leri{s-\frac{\pi \chi}{2}\int_0^\infty d\rho \int_{-\infty}^{+\infty}dp_3\, \dfrac{f'_0(p)\rho^5}{p(p^0s+ikp_3)}}h_{L,R}^{(k)}(0)}{(s^2+k^2)\epsilon_{L,R}(k,s)},
\end{equation}
where, as compared to \cite{Moretti2019}, we introduced the chiral gravitational dielectric functions
\begin{equation}
    \epsilon_{L,R}(k,s)=1\mp \frac{k_{NY} k}{s^2+k^2}
    -\dfrac{\pi\chi}{2(s^2+k^2)}\int d\rho dp_3\, \dfrac{f_0'(p)}{p} \dfrac{s\rho^5}{p^0s+ikp_3},
    \label{dialectric functions NY}
\end{equation}
with $k_{NY}\equiv\alpha \dot{\theta}_B$ having dimensions of momentum, i.e. $[k_{NY}]=[L^{-1}]$. We are interested in the so called \textit{weak damping scenario} \cite{Landau:1946jc,lifshitz1995physical}, where the imaginary part of the frequency is much smaller than its real part, i.e. $|\omega_i|\ll|\omega_r|$ (frequency is related to the Laplace coordinate by $\omega=i s$). In this case, indeed, the characteristic period of oscillation for the gravitational wave is still much smaller than the typical damping time, preventing the formation of transient signals too much rapidly decaying for being detected. The periodic behavior is then encoded in the real part of the dielectric function, and the dispersion relation $\omega_r=\omega_r(k)$ can be obtained by solving
\begin{equation}
    \Re (\epsilon_{L,R})(k,\omega_r)=0.
    \label{real part epsilon}
\end{equation}
Once we have $\omega_r$, we can finally deduce the damping coefficient directly from
\begin{equation}
    \omega_i=-\left.\frac{ \Im(\epsilon_{L,R})}{\frac{\partial\Re (\epsilon_{L,R})}{\partial\omega}}\right|_{\omega=\omega_r}.
    \label{imaginarypart}
\end{equation}
For the J\"uttner-Maxwell background distribution \eqref{max jutt} the dielectric functions are given by:
\begin{align}
    \epsilon_{L,R} (k,\omega_r)=1\mp \frac{k_{NY}}{k(1-v_p^2)}-\frac{n\chi}{4k^2 m^2T^2K_2\leri{x}}\leri{\dfrac{v_p}{1-v_p^2}}^2\int d\rho dp_3 \, \dfrac{\rho^5e^{-\frac{\sqrt{m^2+\rho^2+p_3^2}}{T}}}{p_3^2-\frac{v_p^2 }{1-v_p^2}(m^2+\rho^2)}, \label{dielectric NY jutt}
\end{align}
where $v_p\equiv\frac{\omega_r}{k}$ is the phase velocity. We observe, at this point, that along the integration path of $p_3$, whenever the condition $v_p<1$ holds, one finds a pair of poles located at $p_3=\pm\sqrt{\frac{v_p^2}{1-v_p^2}(m^2+\rho^2)}$. This guarantees the existence of an imaginary part for dielectric functions, which can then be  evaluated by applying the residue theorem. The range of validity of the inequality $v_p<1$ can be established, however, only by solving \eqref{real part epsilon} for $v_p$, and it usually results in phenomenological constraints relating the parameters of the model and the physical properties of the medium (see discussion in \cite{Moretti:2020kpp}).

In order to obtain $\omega_r$, we follow the standard approach of plasma physics \cite{lifshitz1995physical} by expanding the denominator of the integrals in \eqref{dielectric NY jutt} up to the second order in $p_3$. This amounts to assuming  $\frac{p_3}{v_p}\sqrt{\frac{1-v_p^2}{m^2+\rho^2}}\ll 1$, which corresponds to having a phase velocity for the wave much greater than the thermal velocity of the medium. It has to be remarked that this assumption typically holds for material media in weak field regime, such as galactic and Solar System dark matter halos. In fact, in the presence of strong gravity scenarios, characterized by $x\simeq 1$ and corresponding to high density and temperature, this hypothesis is not well grounded and numerical techniques of integration of the dielectric function are usually required (see \cite{Moretti:2020kpp}). The real part of the dielectric function is then given by:
\begin{equation}
\Re (\epsilon_{L,R})=1\pm\frac{k_{NY} }{k(v_p^2-1)}+\frac{
2\omega_0^2}{x^2k^2}\leri{\frac{x}{1-v_p^2}+\frac{\gamma(x)}{v_p^2} },
\label{real epsilon ny}
\end{equation}
where we defined $\gamma(x)\equiv\frac{K_1(x)}{K_2(x)}$ and the proper frequency of the medium $\omega_0^2\equiv\chi n m=\chi \rho$, being $\rho$ the mass density of the medium (see \cite{Montani:2018iqd} for a comparison). Setting $\Re (\epsilon_{L,R})=0$, we finally get the expression for the phase velocity of the left and right polarizations\footnote{We remark that $\Re (\epsilon_{L,R})=0$ admits also a couple of solutions with a minus sign in front of the square root. We disregard these possibilities since they result in the contradiction $\omega_r^2<0$ for every value of $k$. For more detail see the discussion in section \ref{sec3} about the Chern-Simons theory.}, i.e.
\begin{align}
\label{phase velocity L}
    \omega^2_L=\frac{k^2}{2}\leri{1+\frac{k_{NY}}{k}+\frac{2\omega_0^2}{k^2}\frac{x-\gamma(x)}{x^2}+\sqrt{\leri{1+\frac{k_{NY}}{k}+\frac{2\omega_0^2}{k^2}\frac{x-\gamma(x)}{x^2}}^2+\frac{8\omega_0^2}{k^2}\frac{\gamma(x)}{x^2}}},\\
    \label{phase velocity R}
    \omega^2_R=\frac{k^2}{2}\leri{1-\frac{k_{NY}}{k}+\frac{2\omega_0^2}{k^2}\frac{x-\gamma(x)}{x^2}+\sqrt{\leri{1-\frac{k_{NY}}{k}+\frac{2\omega_0^2}{k^2}\frac{x-\gamma(x)}{x^2}}^2+\frac{8\omega_0^2}{k^2}\frac{\gamma(x)}{x^2}}}.
\end{align}
If we now require $v_p^2<1$, we end up with the condition
\begin{equation}
    \frac{1}{k}\leri{\frac{2\omega_0^2}{x k}\mp k_{NY}}<0.
\label{condition damping NY}
\end{equation}
We immediately see that for $k_{NY}=0$, which corresponds to neglect in theory the Nieh-Yan contribution, we completely recover the results of GR (see \cite{Moretti:2020kpp}), since in this case inequality \eqref{condition damping NY} can never be  fulfilled, and gravitational Landau damping cannot take place for tensor modes. When $\alpha\neq 0$, instead, we obtain the following conditions for $k_{NY}>0$ :
\begin{equation}
\text{Left mode: }k<-k_{NY}^0\qquad \text{Right mode: }k>k_{NY}^0,
\label{k allowed NY}
\end{equation}
where
\begin{equation}
    k_{NY}^0\equiv \frac{2\omega_0^2}{k_{NY} x},
\end{equation}
and the case $k_{NY}<0$ is simply obtained by inverting the sign of the inequalities in \eqref{k allowed NY}. We point out that a positive or negative sign of the wavenumber is associated to the direction of propagation of the connected gravitational mode. For instance, having defined a positive direction of the propagation axis, e.g. the direction pointing from the source towards the observer, incoming gravitational waves are characterized by a positive $k$. We note that, in general, damping can only occur for wave numbers satisfying the constraint
\begin{equation}
    |k|> \frac{2\omega_0^2}{|k_{NY}| x}=\frac{2 \chi n T}{|k_{NY}|},
\end{equation}
and we see that the colder and less dense the medium traveled by the gravitational signal, the wider the range of wave numbers affected by the Landau damping. In other words, less relativistic media seem to be more prone to induce kinematic damping in gravitational waves, and this is somehow in contrast with the results of \cite{Moretti:2020kpp}, where relativistic media turned out to be favored.

We calculate now the imaginary part of \eqref{dielectric NY jutt}. Firstly, we note that the Nieh-Yan term does not contribute as long as the parameter $\alpha$ is real. Secondly, integration in $p_3$ can be performed by simply evaluating half the residue in the pole, leading to
\begin{equation}
    \Im(\epsilon_{L,R})=-\frac{\pi\omega_0^2}{4 k^2 m^4 T^2 K_2(x)}\frac{v_p}{(1-v_p^2)^{3/2}}\int_0^{+\infty}d\rho\;\frac{\rho^5 e^{-x\sqrt{\leri{\frac{1}{1-v_p^2}}\leri{1+\frac{\rho^2}{m^2}}}}}{\sqrt{1+\frac{\rho^2}{m^2}}}.
\label{imaginary dielectric}
\end{equation}
It is now convenient to change variable of integration as
\begin{equation}
    \rho=m \sqrt{y^2-1},
\end{equation}
which allows us to recast \eqref{imaginary dielectric} in the following way
\begin{equation}
    \Im(\epsilon_{L,R})=-\frac{\pi\omega_0^2 m^2}{4 k^2T^2 K_2(x)}\frac{v_p}{(1-v_p^2)^{3/2}}\int_1^{+\infty}dy \leri{y^2-1}^2 e^{-\sigma y},
\end{equation}
where $\sigma\equiv x\sqrt{\frac{1}{1-v_p^2}}$, which for $v_p^2<1$ is always positive. The integral can be evaluated by iteratively integrating by parts, resulting in
\begin{equation}
    \int_1^{+\infty}dy \leri{y^2-1}^2 e^{-\sigma y}=\frac{8}{\sigma^3}\leri{1+\frac{3}{\sigma}+\frac{3}{\sigma^2}}e^{-\sigma},
\end{equation}
which inserted in \eqref{imaginary dielectric} gives us
\begin{equation}
    \Im(\epsilon_{L,R})=-\frac{\pi\omega_0^2\; v_p}{k^2 x^3 K_2(x)}\leri{3(1-v_p^2)+3x \leri{1-v_p^2}^{1/2}+x^2 }e^{-x\sqrt{\frac{1}{1-v_p^2}}}.
\end{equation}
Then, from \eqref{imaginarypart} and taking into account \eqref{real epsilon ny}, we obtain the following expression for the damping coefficient:
\begin{equation}\label{imaginary part NY}
    \omega_i=\frac{\pi e^{-\frac{x}{\sqrt{1-v_p^2}}} \left(3 \left(1-v_p^2\right)+3 x \sqrt{1-v_p^2}+x^2\right)}{4 x K_2(x) \left(\frac{x}{ \left(1-v_p^2\right)^2}\leri{1-\frac{k}{k_{NY}^0}}-\frac{ \gamma(x) }{ v_p^4}\right)}k.
\end{equation}
It can be thus verified that for all the wavenumbers which satisfy $v_p^2<1$, as indicated in \eqref{k allowed NY}, the inequality $\omega_i<0$ identically holds, guaranteeing the absence of instabilities due to exponentially growing modes, in agreement with the hypothesis of the phase velocity of the wave being much greater than the average thermal velocity of the particle distribution function. Moreover, the fact that \eqref{imaginary part NY} depends explicitly on $k_{NY}^0$ and indirectly on the chiral state by means of the phase velocities evaluated from \eqref{phase velocity L}-\eqref{phase velocity R}, leads to the interesting result that an initial velocity birefringence can be converted in amplitude birefringence through the Landau mechanism.

\section{Metric Chern-Simons gravity}\label{sec3}
Following the analysis of \cite{Chatzistavrakidis:2021oyp} for the Nieh-Yan case, we can consider an analogous scenario in metric Chern-Simons gravity \cite{Jackiw:2003pm}, looking again at the propagation of gravitational waves on a Minkowski background. In this case the pseudoscalar field $\theta(x)$ is nonminimally coupled to the Pontryagin density in the action, obtained from the contraction of the Riemann tensor with its dual and defined as $\,^{*}RR\equiv\,^{*}R^{\mu\nu\alpha\beta}R_{\nu\mu\alpha\beta}$. From \cite{Nojiri:2019nar}, it is then possible to obtain the equation for the metric perturbation, i.e.
\begin{equation}
    \Box h_{\mu\nu}+4\alpha\dot{\theta}_B(\eta_{\mu k}\eta_{\nu\rho}+\eta_{\mu \rho}\eta_{\nu k})\epsilon^{i j k}\partial_\sigma R\indices{^{\sigma\rho}_{ij}}=-2\chi T_{\mu\nu},
\end{equation}
where we neglected again second order time derivatives for $\theta(x)$ under the assumption of adiabaticity. Moreover, we immediately note that in this case the parity violating contribution carries higher order derivatives than the Nieh-Yan model, more specifically third-order derivatives of the metric perturbation. Now, looking at the purely tensor modes of $h_{\mu\nu}$ and assuming as in section \ref{sec2} a wave travelling in the $z$ direction, we obtain
\begin{align}
    &\Box(h_+-8\alpha \dot{\theta}_B\partial_z h_\times)=-2\chi T_{11}\\
    &\Box(h_\times+8\alpha \dot{\theta}_B\partial_z h_+)=-2\chi T_{12},
\end{align}
which, rewritten in terms of the circularized polarization states, results in
\begin{align}\label{waveqL}
    &\Box(1-8i\alpha \dot{\theta}_B\partial_z)h_L=-2\chi T_L\\\label{waveqR}
    &\Box(1+8i\alpha \dot{\theta}_B\partial_z)h_R=-2\chi T_R.
\end{align}
The dispersion relation describing the propagation in vacuum of the left- and right-handed modes is then simply given by
\begin{equation}
    \mathcal{D}_{L,R}(\omega,k)=\leri{\omega^2-k^2}\leri{1\pm 8\alpha\dot{\theta}_B 
    k}=0,
\end{equation}
which exhibits as general solution the superposition of the wave described by $\omega=\pm k$, with a non-radiative spatial oscillation of wavenumber $k=\pm (8\alpha \dot{\theta}_B)^{-1}$. As a result, birefringence in vacuum is absent and the radiative component of the metric perturbation propagates as in GR. We are interested in describing the behavior of gravitational waves from metric Chern-Simons gravity in the presence of non-collisional matter. As illustrated in the previous section, the stress-energy tensors $T_L$ and $T_R$ in equations \eqref{waveqL} and \eqref{waveqR} can be written in terms of the perturbation in the distribution function of the particles composing the medium due to the presence of gravitational waves. The evolution in time of such probability perturbation is given by the Vlasov equation \eqref{linear vlasov} which, together with the wave equations \eqref{waveqL} and \eqref{waveqR}, constitute a closed system. Solutions are found in the Fourier-Laplace space previously introduced and, for a generic background distribution $f_0(p)$, they read in the Chern-Simons theory as:
\begin{equation}\label{sceqcs}
    h_{L,R}^{(k,s)}=\dfrac{\leri{s\leri{1\pm8\alpha \dot{\theta}_B k}-\frac{\pi \chi}{2}\int_0^\infty d\rho \int_{-\infty}^{+\infty}dp_3\, \dfrac{f'_0(p)\rho^5}{p(p^0s+ikp_3)}}h_{L,R}^{(k)}(0)}{\leri{s^2+k^2} \leri{1\pm8\alpha \dot{\theta}_B k}\epsilon_{L,R}(k,s)},
\end{equation}
with the chiral dielectric functions defined in this case by
\begin{equation}\label{fdiel}
    \epsilon_{L,R}(k,s)=1
    -\dfrac{\pi\chi s\; B_{L,R}(k)}{2\leri{s^2+k^2}}\int d\rho dp_3\, \dfrac{f_0'(p)}{p} \dfrac{
    \rho^5}{p^0s+ikp_3},
\end{equation}
where we introduced the birefringence factor
\begin{equation}
    B_{L,R}(k)\equiv\frac{1}{1\pm 8\alpha\dot{\theta}_B k}\equiv\frac{1}{1\pm\frac{k}{k_{CS}}},
\end{equation}
being $k_{CS}$ the Chern-Simons ``momentum'' defined by analogy with the Nieh-Yan theory, i.e. $k_{CS}\equiv(8\alpha\dot{\theta}_B)^{-1}$.
It is worth mentioning that with respect to the Nieh-Yan model, see \eqref{dialectric functions NY}, the term keeping trace of the parity violating effects appears in front of the integrals involving the distribution function. We expect, therefore, that both the real and the imaginary parts of the dielectric functions be affected by the Pontryagin density. Then, by assuming as in section \ref{sec2} a J\"uttner-Maxwell background distribution \eqref{max jutt}, we derive the real part of $\epsilon_{L,R}$ by expanding the denominator in \eqref{fdiel} under the assumption of phase velocity much greater than thermal velocity of the medium particles and integrating term by term the truncated series obtained, i.e.
\begin{equation}
    \Re(\epsilon_{L,R})=1+\dfrac{2 \omega_0^2 B_{L,R}(k)}{x^2 k^2}\leri{\dfrac{x}{1-\frac{\omega^2}{k^2}}+\dfrac{\gamma(x)}{\frac{\omega^2}{k^2}}}.
\label{real epsilon CS metric}
\end{equation}
The dispersion relations $\omega_r(k)$ are found from $\Re(\epsilon_{L,R})=0$  which, being a quartic equation for the frequency, results in a couple of independent branches of solutions reading 
\begin{align}\label{solplus}
    \omega^2_{L,R}=\frac{k^2}{2}\leri{1+\frac{2\omega_0^2\leri{x-\gamma}B_{L,R}(k)}{x^2k^2}\pm\sqrt{\leri{1+\frac{2\omega_0^2\leri{x-\gamma}B_{L,R}(k)}{x^2k^2}}^2+\frac{8\omega_0^2\gamma B_{L,R}(k)}{x^2 k^2}}},
\end{align}
where the signs corresponding to the left and right polarizations are now encoded in the function $B_{L,R}(k)$. It must be remarked that, in contrast with the findings of the previous section, here we have up to two independent solutions which can satisfy the reality condition $\omega^2>0$, due to the fact that the sign of the function $B_{L,R}(k)$ is not a priori fixed. In the case $B_{L,R}(k)>0$, reality is assured only for the plus sign solution in \eqref{solplus}. Solutions described by the minus sign, instead, do not represent modes that can propagate in the medium, as in this case the condition $\omega^2<0$ identically holds, irrespective of the wavenumber $k$. Moreover, by selecting the plus sign in \eqref{solplus}, we actually obtain gravitational waves with superluminal phase velocity, for which the imaginary part of the dielectric function is strictly null. Consequently, for $B_{L,R}(k)>0$ no damping can occur, and propagation is allowed for $k_{CS}>0$ in the following cases:
\begin{equation}
    \text{Left mode: }k>-k_{CS}\qquad \text{Right mode: }k<k_{CS},
\label{k allowed B pos metric}
\end{equation}
where, again, results for $k_{CS}<0$ can be just derived by inverting the signs of the inequalities. When the opposite case $B_{L,R}(k)<0$ is considered, it is found that the reality condition is fulfilled for both solutions when 
\begin{equation}\label{realitycondition}
    \frac{k^2}{B_{L,R}(k)}+\delta^2=k^2 \leri{1\pm \frac{k}{k_{CS}}}+\delta^2<0,
\end{equation}
where we defined $\delta^2\equiv\frac{2\omega_0^2}{x^2}\leri{\sqrt{x}+\sqrt{\gamma(x)}}^2 $. We stress that the reality of at least one dispersion relation is a necessary condition for the existence of waves supported by the matter: in other words, wavenumbers not satisfying \eqref{realitycondition} and for which $B_{L,R}(k)<0$ can not propagate within the medium.
\\We proceed, thus, by selecting the values of $k$ that guarantee the reality of the dispersion relations. It can be shown that the cubic equations $\pm \frac{k^3}{k_{CS}}+k^2+\delta^2=0$ are characterized by a single real root, reading
\begin{equation}
    k^0_{L,R}=\mp k^0\equiv\mp \frac{\Sigma^2(k_{CS},\delta^2)+\Sigma(k_{CS},\delta^2)+1}{3\Sigma(k_{CS},\delta^2)}\;k_{CS},
\end{equation}
where
\begin{equation}
    \Sigma(k_{CS},\delta^2)\equiv \leri{\frac{2}{\frac{27\delta^2}{k_{CS}^2}+2-\sqrt{\leri{\frac{27\delta^2}{k_{CS}^2}+2}^2-4}}}^\frac{1}{3},
\end{equation}
so that \eqref{realitycondition} is in general satisfied for ranges of wavenumber of the form $|k|>|k^0_{L,R}|$. Furthermore, when reality is verified, it is found that both dispersion relations predict subluminal phase velocity for all $k$. Hence, for wavenumbers in the ranges $|k|>|k^0_{L,R}|$, the propagation throughout the medium is not purely dispersive and Landau damping occurs for $k_{CS}>0$ if
\begin{equation}
    \text{Left mode: }k<k_{L}^0<0\qquad \text{Right mode: }k>k_{R}^0>0,
\label{k allowed B pos metric 2}
\end{equation}
where the case $k_{CS}<0$ is obtained in analogy with the previous section. We remark that being always $|k^0_{L,R}|>|k_{CS}|$, there is an interval of wavenumbers $|k_{CS}|<|k|<|k^0_{L,R}|$ for which both dispersion relations do not exist. For these modes the propagation within the medium is not allowed: they can propagate in vacuum but they are totally reflected when entering the matter medium \cite{stix1992waves}.\\
\noindent
Having described the properties and the range of validity of the dispersion relations, we can proceed with calculating the imaginary part of the dielectric function. We recall that this quantity is non-null only for the wavenumbers indicated in \eqref{k allowed B pos metric 2}, where gravitational modes propagate with subluminal phase velocity. Following the same procedure illustrated in the previous section, we obtain
\begin{equation}
    \Im(\epsilon_{L,R})=-\frac{\pi\omega_0^2 B_{L,R}\; v_p}{k^2 x^3 K_2(x)}\leri{3(1-v_p^2)+3x \leri{1-v_p^2}^{1/2}+x^2 }e^{-x\sqrt{\frac{1}{1-v_p^2}}}.
    \label{imaginary part CS diel}
\end{equation}
Then, by making use of formula \eqref{imaginarypart}, we attain the imaginary part of the frequency, reading

\begin{equation}
    \omega_i=\frac{\pi e^{-\frac{x}{\sqrt{1-v_p^2}}} \left(3 \left(1-v_p^2\right)+3 x \sqrt{1-v_p^2}+x^2\right)}{4 x K_2(x) \left(\frac{x}{ \left(1-v_p^2\right)^2}-\frac{ \gamma(x) }{  v_p^4}\right)}k.
    \label{imaginary part CS}
\end{equation}
We note that the factor $B_{L,R}$ affects the damping rate through the phase velocity $v_p(k)=\frac{\omega_{R,L}(k)}{k}$, so that gravitational waves in metric CSMG show amplitude and frequency birefringence in matter, despite such an effect could be neglected in vacuum. If we now require the absence of instabilities generated by $\omega_i>0$, the following constraint must be fulfilled
\begin{equation}
   \frac{x}{ \left(1-v_p^2\right)^2}-\frac{ \gamma(x) }{  v_p^4}<0.
\end{equation}
This inequality implies a bound on the phase velocity,
\begin{equation}\label{vpbound}
    v_p^2<\dfrac{\sqrt{\gamma}}{\sqrt{x}+\sqrt{\gamma}},
\end{equation}
which can be shown to be satisfied for $k_{CS}>0$ only for the minus solution in \eqref{solplus}, as long as the reality condition \eqref{realitycondition} holds. Conversely, when $k_{CS}<0$ we have that \eqref{vpbound} is satisfied only by the plus-signed dispersion relation. Hence, according to the sign of $k_{CS}$, we discard the solution associated to an amplitude instability within the medium. Then, we consider solely the minus-signed (plus-signed) solution for $k_{CS}>0$ ($k_{CS}<0$), and we deal with a phase velocity within the medium always smaller than
\begin{equation}
    v_p<\bar{v}\equiv \leri{1+\sqrt{\frac{x}{\gamma(x)}}}^{-\frac{1}{2}},
\end{equation}
for any value of the wavenumber as given in \eqref{k allowed B pos metric 2}. It can be checked that the maximum phase velocity $\bar{v}$ is always greater than the mean value of the velocity along the axis of propagation characterizing the background J\"uttner-Maxwell distribution, which turns out to be
\begin{equation}
    v_{th}=\dfrac{1}{\sqrt{3}}\left < \dfrac{p}{p^0}\right >=\dfrac{2}{\sqrt{3}K_2(x)} \dfrac{1+x}{x^2} e^{-x}.
\end{equation}
The ratio $\frac{\bar{v}}{v_{th}}$ is always growing with $x$ and it is greater than $2$ as soon as $x \gtrsim15$, so that a range of allowed phase velocities satisfying the condition of being much greater than the average velocity of particles always exists for large values of $x$.

\section{Metric-affine Chern-Simons gravity}\label{sec4}
As discussed in \cite{Boudet:2022nub,Boudet:2022wmb} Chern-Simons gravity can be also formulated in the context of metric-affine theories, allowing for the presence of a non trivial geometric structure characterized by torsion and nonmetricity \cite{Hehl:1994ue,Iosifidis2021}. In particular, as outlined in \cite{Boudet:2022nub} for a de Sitter background, in this case gravitational waves are affected, like in the metric formulation, by birefringence, which in this framework is induced by the coupling of the torsion perturbations with the metric ones. In order to compare the predictions of the metric-affine case with the results of section \ref{sec2} and section \ref{sec3}, we extend here the analysis of the model discussed in \cite{Boudet:2022nub} to the Minkowski background. Discussing in detail the technical features of the metric-affine CSMG theory goes beyond the aim of this paper, and we refer the reader to the original works \cite{Boudet:2022nub,Boudet:2022wmb} (see also \cite{Iosifidis:2020dck,Jimenez:2022hcz} for the role of parity-odd terms in generic metric-affine theories). For the purpose of this study, it is sufficient to report the system of coupled equations describing the dynamical part of metric and torsion perturbations, which we derive here for a Minkowski background, i.e.
\begin{align}
\delta q_{102} + \frac{\alpha\dot{\theta}_B}{2} \left( 2\alpha\dot{\theta}_B \delta q_{102}'' + \dot{h}_{+}'-\alpha\dot{\theta}_B \dot{h}_{\times}'' \right)&=0,\\
\delta q_{213} + \frac{\alpha\dot{\theta}_B}{2} \left( 2 \alpha\dot{\theta}_B \delta q_{213}'' -h_{+}''+\alpha \dot{\theta}_B h_{\times}'''  \right)&=0,
\end{align}
and
\begin{align}
\ddot{h}_{+}-h_{+}'' +\alpha \dot{\theta}_B \left( h_{\times}''' - \ddot{h}_{\times}' +2 \delta q_{213}'' + 2 \delta \dot{q}_{102}'\right) &= 2\kappa \delta T_{11},\\
\ddot{h}_{\times}-h_{\times}''-2\delta q_{213}' -2\dot{\delta q}_{102}&= 2\kappa\delta T_{12},
\end{align}
where we assumed again \eqref{metric perturbation tt gauge} for the metric perturbation. Regarding the perturbations in the affine sector, we restricted to the tensor modes encoded in the torsion and nonmetricity rank three tensors $q_{\mu\nu\rho}$ and $\Omega_{\mu\nu\rho}$ (see \cite{Iosifidis2021}, for instance). The components $\delta q_{102}$ and $\delta q_{213}$ contain the only independent parts of torsion involved in the dynamics, while the remaining components of torsion and nonmetricity are either algebraically related to these ones, or satisfy harmonic equations in the spatial part, so that they do not propagate as a wave (see app.~\ref{app} for further details).

In vacuum, where the Fourier analysis can be extended also to the time coordinate, torsion can be solved in terms of the circular polarizations as
\begin{align}
    &\delta q_{102}^{(k,\omega)}=\frac{ \alpha\dot{\theta}_B }{2\sqrt{2}}\leri{\frac{ h_L^{(k,\omega)}}{1-\alpha\dot{\theta}_B k}+\frac{ h_R^{(k,\omega)}}{1+\alpha\dot{\theta}_B k}}k\omega,
    \label{q102 fourier}
    \\
    &\delta q_{213}^{(k,\omega)}=-\frac{\alpha\dot{\theta}_B }{2\sqrt{2}}\leri{\frac{ h_L^{(k,\omega)}}{1-\alpha\dot{\theta}_Bk}+\frac{ h_R^{(k,\omega)}}{1+\alpha\dot{\theta}_Bk}}k^2=-\delta q_{102}^{(k,\omega)},
    \label{q213 fourier}
\end{align}
where we used the fact that in vacuum $v_p=1$, as it emerges from the analysis of the metric equations, which simply reduce to:
\begin{align}
    (\omega^2-k^2)h_{L,R}^{(k,\omega)}=0,
    \label{CS affine wave fourier}
\end{align}
reproducing the well-known GR dispersion relation $\omega=\pm k$. Accordingly, in analogy with the metric formulation, no birefringence arises in vacuum. It is interesting to investigate the role of the denominators $(1\pm\alpha\dot{\theta}_B k)^{-1}$ in \eqref{q102 fourier}-\eqref{q213 fourier}, when the evaluation of the torsion perturbation in the space of coordinates is performed through the inverse Fourier transform. Indeed, after having written down the solution for the metric perturbations, resulting in 
\begin{equation}
    h_{L,R}(z,t)=\dfrac{f_{L,R}(z-t)+f_{L,R}(z+t)}{2}+\dfrac{1}{2}\int_{z-t}^{z+t} \, dz' g_{L,R}(z'),
\end{equation}
with
\begin{equation}
    f_{L,R}(z)=h_{L,R}(z,0), \qquad \qquad g_{L,R}(z)=\partial_t h_{L,R}(z,0),
\end{equation}
one obtains the torsion perturbation as
\begin{align}\label{torsionink}
    \delta q_{102}(z,t)=&\dfrac{1}{8\sqrt{2}\pi}\int_{\mathbf{R}}  dk  \leri{\dfrac{k f_L^{(k)}+ig_L^{(k)}}{k-k_*}-\dfrac{k f_R^{(k)}+ig_R^{(k)}}{k+k_*}}ke^{ik(z-t)} \nonumber \\
    -&\dfrac{1}{8\sqrt{2}\pi}\int_{\mathbf{R}} dk \leri{\dfrac{k f_L^{(k)}-ig_L^{(k)}}{k-k_*}-\dfrac{k f_R^{(k)}-ig_R^{(k)}}{k+k_*}}ke^{ik(z+t)},
\end{align}
where $k_*=8k_{CS}$. For the sake of simplicity let us focus on perturbations traveling towards the positive direction of the $z$ axis. The integral in \eqref{torsionink} can be then calculated by exploiting the residue theorem, obtaining
\begin{align}\label{torspertzt}
    \delta q_{102}(z,t)&=\dfrac{i k_*}{8\sqrt{2}} \leri{e^{ik_*(z-t)}\leri{k_*f_L^{(k_*)}+i g_L^{(k_*)}}+e^{-ik_*(z-t)}\leri{k_*f_R^{(-k_*)}+i g_R^{(-k_*)}}}\vartheta\leri{z-t}+\nonumber\\
    &-\dfrac{i k_*}{8\sqrt{2}} \leri{e^{ik_*(z-t)}\leri{k_*f_L^{(k_*)}+i g_L^{(k_*)}}+e^{-ik_*(z-t)}\leri{k_*f_R^{(-k_*)}+i g_R^{(-k_*)}}}\vartheta\leri{t-z}= \nonumber\\
    &=-\dfrac{i k_*}{16\sqrt{2}\pi}\leri{e^{ik_*(z-t)}h_{L(0)}^{(k_*,-k_*)}+e^{-ik_*(z-t)}h_{R(0)}^{(-k_*,-k_*)}}\leri{\vartheta\leri{z-t}-\vartheta\leri{t-z}},
\end{align}
where in the last line we have introduced $\hat{h}_{L,R(0)}^{(k,\omega)}\equiv \hat{h}_{L,R}^{(k,\omega)}\delta\leri{\omega^2-k^2}$, being $\hat{h}_{L,R}^{(k,\omega)}$ the Fourier transform of the circularly polarized gravitational waves. In order to assure the reality of the linearly polarized modes $h_+$ and $h_\times$ the following must hold (see \cite{Isi:2022mbx}):
\begin{equation}
    \overline{h_{L}^{(k,\omega)}}=h_{R}^{(-k,\omega)} \qquad \qquad \text{and} \qquad \qquad \overline{h_{L}^{(k,\omega)}}=h_{R}^{(k,-\omega)}.
\end{equation}
Hence the expression \eqref{torspertzt} giving the torsion perturbation in the coordinate space can be further simplified, resulting in
    \begin{equation}
       \delta q_{102}(z,t) =-\dfrac{i k_*}{8\sqrt{2}\pi}\Re \leri{e^{-ik_*(z-t)}h_{R(0)}^{(k_*,k_*)}}\leri{\vartheta\leri{z-t}-\vartheta\leri{t-z}}. 
    \end{equation}
As opposed to the metric perturbation, which exhibits in principle an arbitrary wavelength, here the torsion perturbations $\delta q_{102}$ and $\delta q_{213}$ appear to be characterized by the critical wavenumber $k_*$. As discussed in app.~\ref{app}, this in turn implies that the perturbations $\delta q_{201}$ and $\delta q_{123}$ are likewise monochromatic, with the remaining modes $\delta q_{110}, \delta q_{202}, \delta q_{223},\delta q_{131}$ propagating instead on the entire spectrum. 

We now consider the propagation within the medium, and in this case the analogous of \eqref{q102 fourier},\eqref{q213 fourier} and \eqref{CS affine wave fourier}, are displayed by\footnote{As in the previous sections we set the initial condition $\dot{h}^{(k)}_{L,R}(0)=0$, which in this case is also consistent with the choice $\delta q_{102}^{(k)}(0)=0$.}:
\begin{align}
    &\delta q_{102}^{(k,s)}=\frac{i \alpha\dot{\theta}_B k}{2\sqrt{2}}\leri{\frac{s h_L^{(k,s)}-h^{(k)}_L(0)}{1-\alpha\dot{\theta}_B k}+\frac{s h_R^{(k,s)}-h^{(k)}_R(0)}{1+\alpha\dot{\theta}_B k}},
    \label{q102 fourier-laplace}
    \\
    &\delta q_{213}^{(k,s)}=-\frac{\alpha\dot{\theta}_B k^2}{2\sqrt{2}}\leri{\frac{ h_L^{(k,s)}}{1-\alpha\dot{\theta}_Bk}+\frac{ h_R^{(k,s)}}{1+\alpha\dot{\theta}_Bk}},
    \label{q213 fourier-laplace}
    \\
    &\frac{(s^2+k^2)h_L^{(k,s)}-s h^{(k)}_L(0)}{1-\alpha\dot{\theta}_B k}=2\chi T_L,
    \label{CS affine wave L}\\
    &\frac{(s^2+k^2)h_R^{(k,s)}-s h^{(k)}_R(0)}{1+\alpha\dot{\theta}_B k}=2\chi T_R.
    \label{CS affine wave R}
\end{align}
The Fourier-Laplace component of the metric perturbation can be rewritten as
\begin{equation}\label{sceqcsaff}
    h_{L,R}^{(k,s)}=\dfrac{\leri{s-\frac{\pi \chi \leri{1\mp\alpha \dot{\theta}_B k}}{2}\int_0^\infty d\rho \int_{-\infty}^{+\infty}dp_3\, \dfrac{f'_0(p)\rho^5}{p(p^0s+ikp_3)}}h_{L,R}^{(k)}(0)}{\leri{s^2+k^2} \epsilon_{L,R}(k,s)},
\end{equation}
where the chiral dielectric functions can be still defined as in \eqref{fdiel}, i.e.
\begin{equation}\label{fdiel CS affine}
    \epsilon_{L,R}(k,s)=1
    -\dfrac{\pi\chi s\; B_{L,R}(k)}{2\leri{s^2+k^2}}\int d\rho dp_3\, \dfrac{f_0'(p)}{p} \dfrac{
    \rho^5}{p^0s+ikp_3},
\end{equation}
but with the birefringence factor given by
\begin{equation}
    B_{L,R}(k)=1\mp\alpha\dot{\theta}_B k=1\mp\frac{k}{8k_{CS}}.
    \label{birefr factor affine}
\end{equation}
The dispersion relations are then obtained as in \eqref{solplus}, with the function $B_{L,R}$ now defined as \eqref{birefr factor affine}. It follows, therefore, that also for the metric-affine case the condition $B_{L,R}>0$ results in the absence of damping for the frequencies described by the plus sign in \eqref{solplus}, corresponding for $k_{CS}>0$ to the wavenumbers
\begin{equation}
    \text{Left mode: }k<8k_{CS}\qquad \text{Right mode: }k>-8k_{CS},\label{Bmaggiore}
\end{equation}
where the conditions for $k_{CS}<0$ are always obtained by inverting the signs of the inequalities.
We note that the metric-affine results can be formally derived from the purely metric ones \eqref{k allowed B pos metric} by reflecting, up to a factor $8$, the regions of the $k$-space where propagation is allowed with respect to the origin in $k=0$.
For $B_{L,R}(k)<0$, the reality condition is instead satisfied for both solutions when \eqref{realitycondition} holds, which in this case assumes the simpler form
\begin{equation}
    k^2+\delta^2\leri{1\mp\frac{k}{8k_{CS}}}>0.
\label{reality condition affine}
\end{equation}
It is easy to check that this inequality is verified for any $k$ when $\delta^2<256k_{CS}^2$. In the opposite case, namely
\begin{equation}
    \delta^2>256k_{CS}^2=\leri{\frac{2}{\alpha\dot{\theta}_B}}^2,
\label{phen constraint}
\end{equation}
it results that the reality of both dispersion relations is achieved for $k_{CS}>0$ when wavenumbers satisfy:
\begin{equation}\label{Bminore}
\text{Left mode: }k<k^0_{-,L} \;\cup\;  k>k^0_{+,L}\qquad \text{Right mode: }
k<k^0_{+,R}\; \cup\;  k>k^0_{-,R},
\end{equation}
where the same considerations hold for the case $k_{CS}<0$ and the wavenumbers defining the boundaries are given by
\begin{align}
    k^0_{\pm,L}&=\frac{\delta^2}{16k_{CS}}\leri{ 1\pm\sqrt{1- \frac{256k_{CS}^2}{\delta^2}}},\\
    k^0_{\pm,R}&=-\frac{\delta^2}{16k_{CS}}\leri{ 1\pm\sqrt{1- \frac{256k_{CS}^2}{\delta^2}}}.
\end{align}
Since $\delta^2$ depends on the thermodynamic properties of the medium, such as temperature and density, relation \eqref{phen constraint} selects matter media in which we observe the appearance of an interval of forbidden wavenumbers. Then, according to the sign of $\Delta\equiv\delta^2-256k_{CS}^2$, and taking into account the conditions that follow from the analysis of $B_{L,R}(k)$, we have that for $k_{CS}>0$ both dispersion relations exist and are characterized by a subluminal phase velocity for $\Delta<0$ when
\begin{align}\label{k allowed CS affine 1}\text{Left mode: }k>8k_{CS}\qquad \text{Right mode: }
k<-8k_{CS},
\end{align}
and for $\Delta>0$ when
\begin{equation}
    \begin{tabular}[h!]{l c c c c c}
\text{Left mode: } & $8k_{CS}<k<k^0_{-,L}$ & & $\cup$ & & $k>k^0_{+,L}$ \\ \\
\text{Right mode: } &
$k<k^0_{+,R}$ & & $\cup$ & &  $k^0_{-,R}<k<-8k_{CS}$
\label{k allowed CS affine 2}
\end{tabular}
\end{equation}
Then, for all modes satisfying $v_p<1$ we calculate a non-zero imaginary part of the frequency, which also in this case has a form identical to \eqref{imaginary part CS}. As explained in the previous section, the bound on the phase velocity \eqref{vpbound} is always satisfied and the sign of $\omega_i$ always results  negative.

\section{Estimates}\label{sec5}
\begin{figure}[h]
    \centering
   \begin{tikzpicture}
   \draw (-4,-0.2) to (-4,0.2)
      node[anchor=south]{$0$}
      ;
    \draw
      (-4,0) to (-2,0)
      node[anchor=north]{$v_p>1$}
      ;
      \draw
      (-2,0) to (0,0)
      ;
      \draw [ultra thick] 
      (0,0) to (2,0)
      node[anchor=north]{$v_p<1$}
      ;
      \draw [ultra thick]
      (2,0) edge[-latex] node[at end, right]{$k$} (4,0)
      ;
      \draw (0,-0.2) to (0,0.2)
      node[anchor=south]{$k^0_{NY}$}
      ;
  \end{tikzpicture}
    \caption{Teleparallel Nieh-Yan model}
    \label{fig1}
\end{figure}
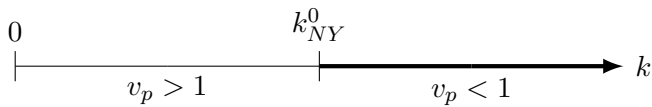

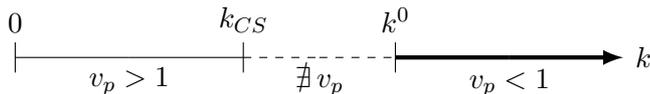
\begin{figure}[h]
    \centering
   \begin{tikzpicture}
   \draw (-4,-0.2) to (-4,0.2)
      node[anchor=south]{$0$}
      ;
    \draw
      (-4,0) to (-2.5,0)
      node[anchor=north]{$v_p>1$}
      ;
      \draw
      (-2.5,0) to (-1,0)
      ;
      \draw [dashed]
      (0,0) to (1,0)
      ;
      \draw [ultra thick]
      (1,0) to (2.5,0)
      node[anchor=north]{$v_p<1$}
      ;
      \draw [ultra thick]
      (2.5,0) edge[-latex] node[at end, right]{$k$} (4,0)
      ;
      \draw (-1,-0.2) to (-1,0.2)
      node[anchor=south]{$k_{CS}$}
      ;
      \draw (1,-0.2) to (1,0.2)
      node[anchor=south]{$k^0$}
      ;
       \draw [dashed] (-1,0) to (0,0)
      node[anchor=north]{$\nexists \, v_p$}
      ;
  \end{tikzpicture}
    \caption{Metric Chern-Simons model}
    \label{fig2}
\end{figure}

 In this section we aim to give a number of quantitative estimates for the predicted effects of amplitude damping on gravitational waves from parity-violating theories of gravity. In all cases, we will assume the concrete scenario of gravitational waves in interaction with a dark matter medium in an environment comparable to the Solar System. Accordingly, as a reference we will take a mass density for the medium $\rho=0.36\pm0.02\, \text{GeV}\,\text{cm}^{-3}$ \cite{sofue2020rotation} and a factor $x\approx 10$. Then, from $\omega_0=\sqrt{\chi \rho}$, we calculate a characteristic frequency of the medium $\omega_0 = 10^{-15}\, \text{Hz} \approx 10^{-23} \, \text{m}^{-1}$. Moreover, we analyze only the case of positive sign for the coupling and for the wavenumber, then retaining just the right-handed polarization (left-handed in the case of metric-affine Chern-Simons gravity), for which damping is expected in the intervals displayed by figs.~\ref{fig1} - \ref{fig4}, where we indicated with normal, thick and dashed lines regions where propagation is, respectively, superluminal, subluminal and forbidden.

\begin{figure}[h]
    \centering
   \begin{tikzpicture}
   \draw (-4,-0.2) to (-4,0.2)
      node[anchor=south]{$0$}
      ;
    \draw
      (-4,0) to (-2,0)
      node[anchor=north]{$v_p>1$}
      ;
      \draw
      (-2,0) to (0,0)
      ;
      \draw [ultra thick] 
      (0,0) to (2,0)
      node[anchor=north]{$v_p<1$}
      ;
      \draw [ultra thick]
      (2,0) edge[-latex] node[at end, right]{$k$} (4,0)
      ;
      \draw (0,-0.2) to (0,0.2)
      node[anchor=south]{$8k_{CS}$}
      ;
  \end{tikzpicture}
    \caption{Metric-affine Chern-Simons model ($\Delta <0$)}
    \label{fig3}
\end{figure}

\begin{figure}[h]
    \centering
   \begin{tikzpicture}
   \draw (-4,-0.2) to (-4,0.2)
      node[anchor=south]{$0$}
      ;
    \draw
      (-4,0) to (-3,0)
      node[anchor=north]{$v_p>1$}
      ;
      \draw
      (-3,0) to (0,0)
      ;
      \draw
      (0,0) to (0.4,0)
      ;
      \draw 
      (1.8,0) to (3,0)
      node[anchor=north]{$v_p<1$}
      ;
      \draw [ultra thick]
      (3,0) edge[-latex] node[at end, right]{$k$} (4,0)
      ;
      \draw (-1.8,-0.2) to (-1.8,0.2)
      node[anchor=south]{$8k_{CS}$}
      ;
      \draw (0.4,-0.2) to (0.4,0.2)
      node[anchor=south]{$k^0_{-}$}
      ;
      \draw (1.8,-0.2) to (1.8,0.2)
      node[anchor=south]{$k^0_{+}$}
      ;
       \draw 
       (-1.5,0) to (-0.6,0)
      node[anchor=north]{$v_p<1$}
      ;
        \draw [dashed] (0.4,0) to (1.2,0)
      node[anchor=north]{$\nexists \, v_p$}
      ;
       \draw [dashed] (1.2,0) to (1.8,0)
       ;
       \draw [ultra thick] (-1.8, 0) to (0.4,0)
       ;
       \draw [ultra thick] (1.8, 0) to (3.8,0)
       ;
  \end{tikzpicture}
    \caption{Metric-affine Chern-Simons model ($\Delta >0$)}
    \label{fig4}
\end{figure}
We remark that the maximum damping is expected for wavenumbers with values close to the coupling of the model considered, i.e. for $k\approx k_{NY}$ in the Nieh-Yan case and $k\approx k_{CS}$ in the Chern-Simons one. As will be shown in the next sections, the typical values of relative absorption in the case of maximum damping are of order $10^{-11}\divisionsymbol 10^{-15}$ (see also \cite{Moretti:2020kpp} for other estimates). The magnitude of the relative absorption is dramatically suppressed as long as wavenumbers outside the region around the coupling value are considered. It is sufficient to consider a wavenumber ten time greater (or smaller) than the coupling value to obtain a completely negligible relative absorption of order smaller than $10^{-30}$.

\subsection{Nieh-Yan}
By considering the tighter estimate for the magnitude of the Nieh-Yan coupling in \cite{Wu:2021ndf,Chatzistavrakidis:2021oyp}, i.e. $k_{NY}=\alpha \dot{\theta}_B \lesssim 5.2 \times 10^{-27}\; \text{m}^{-1}$ we obtain the minimum value for the threshold separating damped from undamped modes, namely $k^0_{NY} \gtrsim 10^{-21}\, \text{m}^{-1}$. Smaller values for the coupling $k_{NY}$ would imply larger values for the threshold $k^0_{NY}$. We observe that the wavelength associated to such transition can live on a spatial scale in principle measurable with ground-based interferometers. A possible test of the theory could be then to search, in the signal detected by a gravitational interferometer, the wavenumber $k^0_{NY}$ identifying the transition between damped and undamped waves. For instance, for the signal collected by current ground-based instruments, which is roughly in the band  $k \in \leri{k^{GB}_{min},k^{GB}_{max}}$ with $k^{GB}_{min}=10^{-8}\, \text{m}^{-1}$ and $k^{GB}_{max}= 10^{-5}\, \text{m}^{-1}$ \cite{2020PhRvD.102f2003B,2021Univ....7..322B}, the analysis could allow to investigate the interval $k_{NY} \in \leri{10^{-42},10^{-39}}\, \text{m}^{-1}$. In other words, if no transition between damped and undamped modes is detected, one can exclude a window of values for $k_{NY}$ lying inside the permitted region provided in \cite{Wu:2021ndf,Chatzistavrakidis:2021oyp}. Considering instead the space interferometer LISA, which is expected to be sensitive to signals in the band $k \in \leri{k^{L}_{min},k^{L}_{max}}$, with $k^{L}_{min}=10^{-14}\, \text{m}^{-1}$ and $k^{L}_{max}=10^{-9}\, \text{m}^{-1}$ \cite{2019CQGra..36j5011R}, the window $k_{NY} \in \leri{10^{-38},10^{-33}}\, \text{m}^{-1}$ would become testable with the observation of a gravitational waves damping from the material medium traversed. However, given that the wavenumbers detectable with either ground-based or space interferometers are much greater than the maximum coupling value $k_{NY}$ provided in \cite{Wu:2021ndf,Chatzistavrakidis:2021oyp}, the expected damping for $k$ inside the intervals $\leri{k^{GB}_{min},k^{GB}_{max}}$ or $\leri{k^{L}_{min},k^{L}_{max}}$ is completely negligible (we calculate values of relative absorption smaller than $10^{-60}$) and a potential transition between damped and undamped modes would be impossible to detect. The resonance returning the maximum damping is expected for $k\approx k_{NY}$, which given the current constraint on the coupling is located in the far large-wavelengths limit. For such a radiation, however, our analysis is not satisfactorily predictive, given that for those wavelengths the approximation of a Minkowski background spacetime is not consistent. Hence, we conclude that a test on this specific model through detection of amplitude modification due to Landau damping is practically impossible.

\subsection{Metric and metric-affine Chern-Simons}
In the case of the metric formulation of Chern-Simons theory we have two distinct effects that can, in principle, be tested with gravitational waves observations, namely the threshold $k_{CS}$ between damped and undamped radiation and the interval of forbidden wavenumbers, whose width is given by $k^0-k_{CS}$.  
We consider the bound on the characteristic length introduced in the Chern-Simons action derived in
\cite{Nakamura:2018yaw}, which in our case reads $k_{CS}\gtrsim 10^{-11}\, \text{m}^{-1}$. For what concerns the interval of forbidden wavenumbers, we proceed by assuming $k_{CS}$ inside the sensitivity curve of ground-based gravitational interferometers, i.e. $k_{CS} \in \leri{k_{min},k_{max}}$, obtaining
\begin{align}
    &k_{CS}=k_{min}  \quad\Longrightarrow\quad k^0-k_{CS}= 10^{-39} \, \text{m}^{-1} \\
    &k_{CS}=k_{max}  \quad\Longrightarrow\quad k^0-k_{CS}= 10^{-42} \, \text{m}^{-1}.
\end{align}
Given the chosen values for $\omega_0$ and $x$, we have the appearance of an extremely narrow window of not-allowed modes which is realistically not detectable. However, the width of this interval depends quadratically on the proper frequency $\omega_0$, hence if the signal travels in a denser environment we expect a considerable increase of this window. Now we are interested in giving an estimate of the amplitude damping detectable for wavenumbers around $k_{CS}$. For instance, assuming $k=k^0$ and $ k_{CS}=10^{-7}\, \text{m}^{-1}$, we obtain from \eqref{imaginary part CS} an imaginary part of the frequency $\omega_i \approx -10^{-15}\, \text{Hz}$. We can calculate the total time of interaction between waves and medium by dividing the proper length of the Solar System, which we take as $L=6 \cdot 10^{12} \, \text{m}$, by the phase velocity associated to the considered wavenumber times the relativistic factor $\gamma=\leri{1-v_p^2}^{-\frac{1}{2}}$, giving account of the contraction of the path length in the frame comoving with the wave phases. We obtain $\Delta T = 3.5 \cdot 10^4 \, \text{s}$. Normalizing to $1$ the amplitude of the waves before the interaction and defining $\mathcal{A}$ as the amplitude after the interaction takes place, we expect a relative absorption of
\begin{equation}\label{reldamping}
    1-\mathcal{A}= \mathcal{O}\leri{10^{-11}}.
\end{equation}
Regarding the case of the metric-affine formulation of Chern-Simons gravity we have shown that the appearance of the window of forbidden wavenumbers is enabled when inequality \eqref{phen constraint} is verified. However, taking into account the bound on $k_{CS}$ and the typical values for $\omega_0$, we observe that $\delta \ll k_{CS}$ for every realistic astrophysical scenario. Hence, in this case, the domain of allowed modes inside the medium is always continuous and a test on the theory can be carried out only by investigating the threshold effect. Given the similarities with the metric case we expect, for wavenumbers around $k_{CS}$, an amplitude damping comparable with that calculated in \eqref{reldamping}.  
\section{Summary and discussion}\label{sec6}
In this work we have analyzed the behavior of gravitational waves from parity-violating theories of gravity propagating in a medium of collisionless particles. As opposed to the case of GR, we have found that, with great generality, tensorial perturbations can indeed be endowed with a subluminal phase velocity when propagating in matter, meeting the necessary condition for a non-vanishing imaginary part of the frequency, thus for a modification in their amplitude. This occurrence is due to the introduction in the action and in the wave equations of a characteristic length scale $\ell$, proportional to $(\alpha \dot{\theta}_B)^{-1}$ in the teleparallel Nieh-Yan case and to $\alpha \dot{\theta}_B$ in both the metric and metric-affine formulations of Chern-Simons gravity, which acts as a threshold separating two distinct regimes for the wave physics within the medium. We have found that, in general, gravitational radiation with wavelength exceeding the bound determined by $\ell$ will be characterized by superluminal phase velocity, so that the amplitude of these waves will remain constant throughout the propagation in matter. On the contrary, for short wavelengths, an amplitude modification due to a mechanism of kinematic energy exchange between the waves and the medium is expected, in full analogy with the phenomenon of Landau damping that arises in electromagnetic plasmas. In our view, one of the most surprising findings in this work is certainly the presence, in the case of Chern-Simons theory, of windows of forbidden wavenumbers, representing not-allowed modes for self-sustained Langmuir excitations within the gravitational plasma. Specifically, we have shown that this interval is always present in the metric case, whereas in the metric-affine formulation its appearance is triggered by the fulfillment of a condition between theory parameters and physical quantities characterizing the medium, i.e. inequality \eqref{phen constraint}. Although this fact is of great theoretical importance, it turns out that these windows are extremely narrow, so that the impact of this particular finding on observations is realistically weak. In spite of that, the threshold effect due to the coupling $k_{CS}$ remains in principle testable with gravitational waves observations, due to the fact that the resonance giving the maximum damping for $k\approx k_{CS}$ is inside the range of wavenumbers in which present-day instruments operate, even though the magnitude of the effect is rather small, as indicated in \eqref{reldamping}, and it is therefore well below the sensitivity of current observations. Unfortunately this is not the case when the teleparallel Nieh-Yan model is considered: here the coupling $k_{NY}$ is constrained  to be far smaller than the smallest wavenumber detectable with ground-based interferometers, hence the expected amplitude modification for measurable wavelengths is negligible, preventing the possibility of a test through gravitational waves damping. We note, moreover, that for a given wavenumber $k$, the Landau damping always involves only one polarization state, so that it can be interpreted as an amplitude birefringence effect induced by an initial velocity birefringence.

An intriguing result in the metric-affine formulation of CSMG is the dynamical propagation of the torsion. In particular we have shown that, in vacuum, specific components of the tensor $q_{\mu\nu\rho}$ are monochromatic with fixed wavelength determined by the parity-violating coefficient of the theory. This property is conserved when the propagation in matter is addressed, as it can be appreciated by looking at the solutions \eqref{q102 fourier-laplace} and \eqref{q213 fourier-laplace}, still characterized by the same couple of poles in $k=\pm k_*$ which again selects a fixed wavelength when the inverse Fourier transform is performed. Nevertheless, several other tensor torsion components do not share this feature as they exhibit a full spectrum of wavelengths, given their dependence on the metric perturbation and the absence of poles in $k$. Without entering in the long standing debate about the observable effects associated to a non trivial metric-affine geometry, here we suggest that in full analogy with the metric formulation of CSMG, the radiative behavior of torsion could result in additional contribution to the stress energy pseudotensor of gravitational waves \cite{Guarrera:2007tu,Stein:2010pn,Bhattacharyya:2018hsj}. These can in principle affect the energy flux emitted in physical processes like binaries ring down, whose analysis will be the object of a following work.

\acknowledgments
The work of F.B is supported by the postdoctoral grant CIAPOS/2021/169. The work of F. M. is supported by the Della Riccia foundation grant for the year 2022. This work is supported by the Spanish Grant PID2020-116567GB-C21 funded by MCIN/AEI/10.13039/501100011033, the project PROMETEO/2020/079 (Generalitat Valenciana), and by the European Union's Horizon 2020 research and innovation programme under the H2020-MSCA-RISE-2017 Grant No. FunFiCO-777740.

\appendix
\section{Torsion and nonmetricity components}\label{app}
We report here the solutions for the rank-3 tensors perturbations of torsion and nonmetricity. Regarding the torsion rank-3 tensor $\delta q_{\mu\nu\rho}$, some of its components propagate as waves, and are ultimately related to the metric perturbations and to the two components $\delta q_{102}$ and $\delta q_{213}$ appearing in the equations of section \ref{sec4}. They are
\begin{align}
    \delta q_{201} &= \delta q_{102},\\
    \delta q_{123} &= \delta q_{213},\\
    \delta q_{110} &= \delta q_{202} = \alpha\dot{\theta}_B \left( \delta q_{102}' -\frac{1}{2} \dot{h}'_{12}  \right),\\
    \delta q_{223} &= \delta q_{131} = \alpha\dot{\theta}_B \left( \delta q_{213}' +\frac{1}{2} h''_{12}  \right).
\end{align}
The following components instead are not propagating:
\begin{equation}
     \delta q_{010}, \quad \delta q_{313}, \quad \delta q_{020}, \quad \delta q_{323}, \quad  \delta q_{031}, \quad \delta q_{301} \quad  \delta q_{032}, \quad \delta q_{302}.
\end{equation}
Rather, they all satisfy the same harmonic oscillator equation, reading
\begin{equation}\label{eq harmonic osc}
    u(t,z) + \frac{\alpha^2 \dot{\theta}^2}{4} u''(t,z) = 0,
\end{equation}
whose solution reads
\begin{equation}
    u(t,z) = C_1(t) \cos\left( \frac{2z}{\alpha\dot{\theta}_B} \right)+C_2(t) \sin\left( \frac{2z}{\alpha\dot{\theta}_B} \right),
\end{equation}
where $C_1$ and $C_2$ are arbitrary functions of time. The nonvanishing components of the nonmetricity rank-3 tensor $\delta\Omega_{\mu\nu\rho}$ are 
\begin{align}
    \delta \Omega_{n00}, \quad \delta \Omega_{n11}, \quad \delta \Omega_{n22}, \quad \delta \Omega_{n03}, \quad \delta \Omega_{n12}, \quad \delta \Omega_{n33},
\end{align}
with $n=1,2$ and
\begin{align}
     \delta \Omega_{n01}, \quad \delta \Omega_{n02}, \quad \delta \Omega_{n13}, \quad \delta \Omega_{n23},
\end{align}
with $n=0,3$ and they all satisfy equation \eqref{eq harmonic osc} as well.

\bibliographystyle{JHEP}
\bibliography{references}

\providecommand{\href}[2]{#2}\begingroup\raggedright\begin{thebibliography}{100}

\bibitem{DeFelice:2010aj}
A.~De~Felice and S.~Tsujikawa, \emph{{f(R) theories}},
  \href{https://doi.org/10.12942/lrr-2010-3}{\emph{Living Rev. Rel.} {\bfseries
  13} (2010) 3} [\href{https://arxiv.org/abs/1002.4928}{{\ttfamily
  1002.4928}}].

\bibitem{Nojiri:2010wj}
S.~Nojiri and S.D.~Odintsov, \emph{{Unified cosmic history in modified gravity:
  from F(R) theory to Lorentz non-invariant models}},
  \href{https://doi.org/10.1016/j.physrep.2011.04.001}{\emph{Phys. Rept.}
  {\bfseries 505} (2011) 59} [\href{https://arxiv.org/abs/1011.0544}{{\ttfamily
  1011.0544}}].

\bibitem{Olmo:2011uz}
G.J.~Olmo, \emph{{Palatini Approach to Modified Gravity: f(R) Theories and
  Beyond}}, \href{https://doi.org/10.1142/S0218271811018925}{\emph{Int. J. Mod.
  Phys. D} {\bfseries 20} (2011) 413}
  [\href{https://arxiv.org/abs/1101.3864}{{\ttfamily 1101.3864}}].

\bibitem{Cai:2015emx}
Y.-F.~Cai, S.~Capozziello, M.~De~Laurentis and E.N.~Saridakis, \emph{{f(T)
  teleparallel gravity and cosmology}},
  \href{https://doi.org/10.1088/0034-4885/79/10/106901}{\emph{Rept. Prog.
  Phys.} {\bfseries 79} (2016) 106901}
  [\href{https://arxiv.org/abs/1511.07586}{{\ttfamily 1511.07586}}].

\bibitem{NOJIRI20171}
S.~Nojiri, S.~Odintsov and V.~Oikonomou, \emph{Modified gravity theories on a
  nutshell: Inflation, bounce and late-time evolution},
  \href{https://doi.org/https://doi.org/10.1016/j.physrep.2017.06.001}{\emph{Physics
  Reports} {\bfseries 692} (2017) 1}.

\bibitem{Krssak:2018ywd}
M.~Krssak, R.J.~van~den Hoogen, J.G.~Pereira, C.G.~B\"ohmer and A.A.~Coley,
  \emph{{Teleparallel theories of gravity: illuminating a fully invariant
  approach}}, \href{https://doi.org/10.1088/1361-6382/ab2e1f}{\emph{Class.
  Quant. Grav.} {\bfseries 36} (2019) 183001}
  [\href{https://arxiv.org/abs/1810.12932}{{\ttfamily 1810.12932}}].

\bibitem{Olmo:2019flu}
G.J.~Olmo, D.~Rubiera-Garcia and A.~Wojnar, \emph{{Stellar structure models in
  modified theories of gravity: Lessons and challenges}},
  \href{https://doi.org/10.1016/j.physrep.2020.07.001}{\emph{Phys. Rept.}
  {\bfseries 876} (2020) 1} [\href{https://arxiv.org/abs/1912.05202}{{\ttfamily
  1912.05202}}].

\bibitem{Cabral:2020fax}
F.~Cabral, F.S.N.~Lobo and D.~Rubiera-Garcia, \emph{{Fundamental Symmetries and
  Spacetime Geometries in Gauge Theories of Gravity\textemdash{}Prospects for
  Unified Field Theories}},
  \href{https://doi.org/10.3390/universe6120238}{\emph{Universe} {\bfseries 6}
  (2020) 238} [\href{https://arxiv.org/abs/2012.06356}{{\ttfamily
  2012.06356}}].

\bibitem{Harko:2020ibn}
T.~Harko and F.S.N.~Lobo, \emph{{Beyond Einstein\textquoteright{}s General
  Relativity: Hybrid metric-Palatini gravity and curvature-matter couplings}},
  \href{https://doi.org/10.1142/S0218271820300086}{\emph{Int. J. Mod. Phys. D}
  {\bfseries 29} (2020) 2030008}
  [\href{https://arxiv.org/abs/2007.15345}{{\ttfamily 2007.15345}}].

\bibitem{Capozziello:2022lic}
S.~Capozziello and F.~Bajardi, \emph{{Non-Local Gravity Cosmology: an
  Overview}},  \href{https://arxiv.org/abs/2201.04512}{{\ttfamily 2201.04512}}.

\bibitem{Fernandes:2022zrq}
P.G.S.~Fernandes, P.~Carrilho, T.~Clifton and D.J.~Mulryne, \emph{{The 4D
  Einstein\textendash{}Gauss\textendash{}Bonnet theory of gravity: a review}},
  \href{https://doi.org/10.1088/1361-6382/ac500a}{\emph{Class. Quant. Grav.}
  {\bfseries 39} (2022) 063001}
  [\href{https://arxiv.org/abs/2202.13908}{{\ttfamily 2202.13908}}].

\bibitem{Yan:2019gbw}
S.-F.~Yan, P.~Zhang, J.-W.~Chen, X.-Z.~Zhang, Y.-F.~Cai and E.N.~Saridakis,
  \emph{{Interpreting cosmological tensions from the effective field theory of
  torsional gravity}},
  \href{https://doi.org/10.1103/PhysRevD.101.121301}{\emph{Phys. Rev. D}
  {\bfseries 101} (2020) 121301}
  [\href{https://arxiv.org/abs/1909.06388}{{\ttfamily 1909.06388}}].

\bibitem{SolaPeracaula:2020vpg}
J.~Sol\`a~Peracaula, A.~G\'omez-Valent, J.~de~Cruz~P\'erez and
  C.~Moreno-Pulido, \emph{{Brans\textendash{}Dicke cosmology with a
  $\Lambda$-term: a possible solution to $\Lambda$CDM tensions}},
  \href{https://doi.org/10.1088/1361-6382/abbc43}{\emph{Class. Quant. Grav.}
  {\bfseries 37} (2020) 245003}
  [\href{https://arxiv.org/abs/2006.04273}{{\ttfamily 2006.04273}}].

\bibitem{Schoneberg:2021qvd}
N.~Sch\"oneberg, G.~Franco~Abell\'an, A.~P\'erez~S\'anchez, S.J.~Witte,
  V.~Poulin and J.~Lesgourgues, \emph{{The H0 Olympics: A fair ranking of
  proposed models}},
  \href{https://doi.org/10.1016/j.physrep.2022.07.001}{\emph{Phys. Rept.}
  {\bfseries 984} (2022) 1} [\href{https://arxiv.org/abs/2107.10291}{{\ttfamily
  2107.10291}}].

\bibitem{DiValentino:2021izs}
E.~Di~Valentino, O.~Mena, S.~Pan, L.~Visinelli, W.~Yang, A.~Melchiorri et~al.,
  \emph{{In the realm of the Hubble tension\textemdash{}a review of
  solutions}}, \href{https://doi.org/10.1088/1361-6382/ac086d}{\emph{Class.
  Quant. Grav.} {\bfseries 38} (2021) 153001}
  [\href{https://arxiv.org/abs/2103.01183}{{\ttfamily 2103.01183}}].

\bibitem{Dainotti:2021pqg}
M.G.~Dainotti, B.~De~Simone, T.~Schiavone, G.~Montani, E.~Rinaldi and
  G.~Lambiase, \emph{{On the Hubble constant tension in the SNe Ia Pantheon
  sample}}, \href{https://doi.org/10.3847/1538-4357/abeb73}{\emph{Astrophys.
  J.} {\bfseries 912} (2021) 150}
  [\href{https://arxiv.org/abs/2103.02117}{{\ttfamily 2103.02117}}].

\bibitem{Ng:2020qpk}
K.K.Y.~Ng, S.~Vitale, W.M.~Farr and C.L.~Rodriguez, \emph{{Probing multiple
  populations of compact binaries with third-generation gravitational-wave
  detectors}}, \href{https://doi.org/10.3847/2041-8213/abf8be}{\emph{Astrophys.
  J. Lett.} {\bfseries 913} (2021) L5}
  [\href{https://arxiv.org/abs/2012.09876}{{\ttfamily 2012.09876}}].

\bibitem{Takeda:2019gwk}
H.~Takeda, A.~Nishizawa, K.~Nagano, Y.~Michimura, K.~Komori, M.~Ando et~al.,
  \emph{{Prospects for gravitational-wave polarization tests from compact
  binary mergers with future ground-based detectors}},
  \href{https://doi.org/10.1103/PhysRevD.100.042001}{\emph{Phys. Rev. D}
  {\bfseries 100} (2019) 042001}
  [\href{https://arxiv.org/abs/1904.09989}{{\ttfamily 1904.09989}}].

\bibitem{Isi:2022mbx}
M.~Isi, \emph{{Parametrizing gravitational-wave polarizations}},
  \href{https://arxiv.org/abs/2208.03372}{{\ttfamily 2208.03372}}.

\bibitem{Hawking:1966qi}
S.W.~Hawking, \emph{{Perturbations of an expanding universe}},
  \href{https://doi.org/10.1086/148793}{\emph{Astrophys. J.} {\bfseries 145}
  (1966) 544}.

\bibitem{Madore}
J.~Madore, \emph{The absorption of gravitational radiation by a dissipative
  fluid}, \href{https://doi.org/10.1007/BF01645508}{\emph{Communications in
  Mathematical Physics} {\bfseries 30} (1973) }.

\bibitem{Madore:1972ww}
J.~Madore, \emph{{The dispersion of gravitational waves}},
  \href{https://doi.org/10.1007/BF01645516}{\emph{Commun. Math. Phys.}
  {\bfseries 27} (1972) 291}.

\bibitem{Prasanna:1999pn}
A.R.~Prasanna, \emph{{Propagation of gravitational waves through a dispersive
  medium}}, \href{https://doi.org/10.1016/S0375-9601(99)00313-8}{\emph{Phys.
  Lett. A} {\bfseries 257} (1999) 120}.

\bibitem{Anile}
A.M.~{Anile} and V.~{Pirronello}, \emph{{High-frequency gravitational waves in
  a dissipative fluid.}}, \href{https://doi.org/10.1007/BF02748651}{\emph{Nuovo
  Cimento B Serie} {\bfseries 48} (1978) 90}.

\bibitem{1978SvA....22..528Z}
A.V.~{Zakharov}, \emph{{A kinetic theory for the growth of perturbations in an
  isotropic cosmological model, and the ultrarelativistic limit}}, {\emph{Sov.
  Astron.} {\bfseries 22} (1978) 528}.

\bibitem{Weinberg:2003ur}
S.~Weinberg, \emph{{Damping of tensor modes in cosmology}},
  \href{https://doi.org/10.1103/PhysRevD.69.023503}{\emph{Phys. Rev. D}
  {\bfseries 69} (2004) 023503}
  [\href{https://arxiv.org/abs/astro-ph/0306304}{{\ttfamily
  astro-ph/0306304}}].

\bibitem{Flauger:2017ged}
R.~Flauger and S.~Weinberg, \emph{{Gravitational Waves in Cold Dark Matter}},
  \href{https://doi.org/10.1103/PhysRevD.97.123506}{\emph{Phys. Rev. D}
  {\bfseries 97} (2018) 123506}
  [\href{https://arxiv.org/abs/1801.00386}{{\ttfamily 1801.00386}}].

\bibitem{Lattanzi:2005xb}
M.~Lattanzi and G.~Montani, \emph{{On the interaction between thermalized
  neutrinos and cosmological gravitational waves above the electroweak
  unification scale}},
  \href{https://doi.org/10.1142/S0217732305018827}{\emph{Mod. Phys. Lett. A}
  {\bfseries 20} (2005) 2607}
  [\href{https://arxiv.org/abs/astro-ph/0508364}{{\ttfamily
  astro-ph/0508364}}].

\bibitem{Lattanzi:2010gn}
M.~Lattanzi, R.~Benini and G.~Montani, \emph{{A possible signature of cosmic
  neutrino decoupling in the nHz region of the spectrum of primordial
  gravitational waves}},
  \href{https://doi.org/10.1088/0264-9381/27/19/194008}{\emph{Class. Quant.
  Grav.} {\bfseries 27} (2010) 194008}
  [\href{https://arxiv.org/abs/1010.3849}{{\ttfamily 1010.3849}}].

\bibitem{Benini:2010zz}
R.~Benini, M.~Lattanzi and G.~Montani, \emph{{Signatures of the neutrino
  thermal history in the spectrum of primordial gravitational waves}},
  \href{https://doi.org/10.1007/s10714-010-0994-4}{\emph{Gen. Rel. Grav.}
  {\bfseries 43} (2011) 945} [\href{https://arxiv.org/abs/1009.6110}{{\ttfamily
  1009.6110}}].

\bibitem{Krause:1994ar}
D.~Krause, H.T.~Kloor and E.~Fischbach, \emph{{Multipole radiation from massive
  fields: Application to binary pulsar systems}},
  \href{https://doi.org/10.1103/PhysRevD.49.6892}{\emph{Phys. Rev. D}
  {\bfseries 49} (1994) 6892}.

\bibitem{Zhang:2017srh}
X.~Zhang, T.~Liu and W.~Zhao, \emph{{Gravitational radiation from compact
  binary systems in screened modified gravity}},
  \href{https://doi.org/10.1103/PhysRevD.95.104027}{\emph{Phys. Rev. D}
  {\bfseries 95} (2017) 104027}
  [\href{https://arxiv.org/abs/1702.08752}{{\ttfamily 1702.08752}}].

\bibitem{Brito:2015oca}
R.~Brito, V.~Cardoso and P.~Pani, \emph{{Superradiance}: {New Frontiers in
  Black Hole Physics}},
  \href{https://doi.org/10.1007/978-3-319-19000-6}{\emph{Lect. Notes Phys.}
  {\bfseries 906} (2015) pp.1}
  [\href{https://arxiv.org/abs/1501.06570}{{\ttfamily 1501.06570}}].

\bibitem{LIGOScientific:2018dkp}
{\scshape LIGO Scientific, Virgo} collaboration, \emph{{Tests of General
  Relativity with GW170817}},
  \href{https://doi.org/10.1103/PhysRevLett.123.011102}{\emph{Phys. Rev. Lett.}
  {\bfseries 123} (2019) 011102}
  [\href{https://arxiv.org/abs/1811.00364}{{\ttfamily 1811.00364}}].

\bibitem{Wagle:2018tyk}
P.~Wagle, N.~Yunes, D.~Garfinkle and L.~Bieri, \emph{{Hair loss in parity
  violating gravity}},
  \href{https://doi.org/10.1088/1361-6382/ab0eed}{\emph{Class. Quant. Grav.}
  {\bfseries 36} (2019) 115004}
  [\href{https://arxiv.org/abs/1812.05646}{{\ttfamily 1812.05646}}].

\bibitem{Chesters:1973wan}
D.~Chesters, \emph{{Dispersion of Gravitational Waves by a Collisionless Gas}},
  \href{https://doi.org/10.1103/PhysRevD.7.2863}{\emph{Phys. Rev. D} {\bfseries
  7} (1973) 2863}.

\bibitem{PhysRevD.13.2724}
E.~Asseo, D.~Gerbal, J.~Heyvaerts and M.~Signore, \emph{General-relativistic
  kinetic theory of waves in a massive particle medium},
  \href{https://doi.org/10.1103/PhysRevD.13.2724}{\emph{Phys. Rev. D}
  {\bfseries 13} (1976) 2724}.

\bibitem{Gayer:1979ff}
S.~Gayer and C.~Kennel, \emph{{Possibility of Landau damping of gravitational
  waves}}, \href{https://doi.org/10.1103/PhysRevD.19.1070}{\emph{Phys. Rev. D}
  {\bfseries 19} (1979) 1070}.

\bibitem{Baym:2017xvh}
G.~Baym, S.P.~Patil and C.~Pethick, \emph{{Damping of gravitational waves by
  matter}}, \href{https://doi.org/10.1103/PhysRevD.96.084033}{\emph{Phys. Rev.
  D} {\bfseries 96} (2017) 084033}
  [\href{https://arxiv.org/abs/1707.05192}{{\ttfamily 1707.05192}}].

\bibitem{Garg:2021baw}
D.~Garg and I.Y.~Dodin, \emph{{Gauge-invariant gravitational waves in matter
  beyond linearized gravity}},
  \href{https://arxiv.org/abs/2106.05062}{{\ttfamily 2106.05062}}.

\bibitem{Garg:2022wdm}
D.~Garg and I.Y.~Dodin, \emph{{Gravitational wave modes in matter}},
  \href{https://doi.org/10.1088/1475-7516/2022/08/017}{\emph{JCAP} {\bfseries
  08} (2022) 017} [\href{https://arxiv.org/abs/2204.09095}{{\ttfamily
  2204.09095}}].

\bibitem{Moretti:2020kpp}
F.~Moretti, F.~Bombacigno and G.~Montani, \emph{{Gravitational Landau Damping
  for massive scalar modes}},
  \href{https://doi.org/10.1140/epjc/s10052-020-08769-z}{\emph{Eur. Phys. J. C}
  {\bfseries 80} (2020) 1203}
  [\href{https://arxiv.org/abs/2005.08010}{{\ttfamily 2005.08010}}].

\bibitem{Moretti:2021ljj}
F.~Moretti, F.~Bombacigno and G.~Montani, \emph{{The Role of Longitudinal
  Polarizations in Horndeski and Macroscopic Gravity: Introducing Gravitational
  Plasmas}}, \href{https://doi.org/10.3390/universe7120496}{\emph{Universe}
  {\bfseries 7} (2021) 496} [\href{https://arxiv.org/abs/2111.11342}{{\ttfamily
  2111.11342}}].

\bibitem{Lombriser:2015sxa}
L.~Lombriser and A.~Taylor, \emph{{Breaking a Dark Degeneracy with
  Gravitational Waves}},
  \href{https://doi.org/10.1088/1475-7516/2016/03/031}{\emph{JCAP} {\bfseries
  03} (2016) 031} [\href{https://arxiv.org/abs/1509.08458}{{\ttfamily
  1509.08458}}].

\bibitem{Kobayashi:2019hrl}
T.~Kobayashi, \emph{{Horndeski theory and beyond: a review}},
  \href{https://doi.org/10.1088/1361-6633/ab2429}{\emph{Rept. Prog. Phys.}
  {\bfseries 82} (2019) 086901}
  [\href{https://arxiv.org/abs/1901.07183}{{\ttfamily 1901.07183}}].

\bibitem{Crisostomi:2016czh}
M.~Crisostomi, K.~Koyama and G.~Tasinato, \emph{{Extended Scalar-Tensor
  Theories of Gravity}},
  \href{https://doi.org/10.1088/1475-7516/2016/04/044}{\emph{JCAP} {\bfseries
  04} (2016) 044} [\href{https://arxiv.org/abs/1602.03119}{{\ttfamily
  1602.03119}}].

\bibitem{bombacigno2023}
F.~Bombacigno, F.~Moretti and G.J.~Olmo, \emph{{Gravitational waves in
  $f(R,\mathcal{L}_m)$ theories: interaction with matter}},
  \href{https://arxiv.org/abs/to appear}{{\ttfamily to appear}}.

\bibitem{Harko:2010mv}
T.~Harko and F.S.N.~Lobo, \emph{{f(R,$L_{m}$) gravity}},
  \href{https://doi.org/10.1140/epjc/s10052-010-1467-3}{\emph{Eur. Phys. J. C}
  {\bfseries 70} (2010) 373} [\href{https://arxiv.org/abs/1008.4193}{{\ttfamily
  1008.4193}}].

\bibitem{Harko:2011kv}
T.~Harko, F.S.N.~Lobo, S.~Nojiri and S.D.~Odintsov, \emph{{$f(R,T)$ gravity}},
  \href{https://doi.org/10.1103/PhysRevD.84.024020}{\emph{Phys. Rev. D}
  {\bfseries 84} (2011) 024020}
  [\href{https://arxiv.org/abs/1104.2669}{{\ttfamily 1104.2669}}].

\bibitem{Barrientos:2018cnx}
E.~Barrientos, F.S.N.~Lobo, S.~Mendoza, G.J.~Olmo and D.~Rubiera-Garcia,
  \emph{{Metric-affine f(R,T) theories of gravity and their applications}},
  \href{https://doi.org/10.1103/PhysRevD.97.104041}{\emph{Phys. Rev. D}
  {\bfseries 97} (2018) 104041}
  [\href{https://arxiv.org/abs/1803.05525}{{\ttfamily 1803.05525}}].

\bibitem{Conroy:2019ibo}
A.~Conroy and T.~Koivisto, \emph{{Parity-Violating Gravity and GW170817 in
  Non-Riemannian Cosmology}},
  \href{https://doi.org/10.1088/1475-7516/2019/12/016}{\emph{JCAP} {\bfseries
  12} (2019) 016} [\href{https://arxiv.org/abs/1908.04313}{{\ttfamily
  1908.04313}}].

\bibitem{Qiao:2019wsh}
J.~Qiao, T.~Zhu, W.~Zhao and A.~Wang, \emph{{Waveform of gravitational waves in
  the ghost-free parity-violating gravities}},
  \href{https://doi.org/10.1103/PhysRevD.100.124058}{\emph{Phys. Rev. D}
  {\bfseries 100} (2019) 124058}
  [\href{https://arxiv.org/abs/1909.03815}{{\ttfamily 1909.03815}}].

\bibitem{Zhao:2019xmm}
W.~Zhao, T.~Zhu, J.~Qiao and A.~Wang, \emph{{Waveform of gravitational waves in
  the general parity-violating gravities}},
  \href{https://doi.org/10.1103/PhysRevD.101.024002}{\emph{Phys. Rev. D}
  {\bfseries 101} (2020) 024002}
  [\href{https://arxiv.org/abs/1909.10887}{{\ttfamily 1909.10887}}].

\bibitem{Chatzistavrakidis:2021oyp}
A.~Chatzistavrakidis, G.~Karagiannis, G.~Manolakos and P.~Schupp, \emph{{Axion
  gravitodynamics, Lense-Thirring effect, and gravitational waves}},
  \href{https://doi.org/10.1103/PhysRevD.105.104029}{\emph{Phys. Rev. D}
  {\bfseries 105} (2022) 104029}
  [\href{https://arxiv.org/abs/2111.04388}{{\ttfamily 2111.04388}}].

\bibitem{Hohmann:2022wrk}
M.~Hohmann and C.~Pfeifer, \emph{{Gravitational wave birefringence in spatially
  curved teleparallel cosmology}},
  \href{https://doi.org/10.1016/j.physletb.2022.137437}{\emph{Phys. Lett. B}
  {\bfseries 834} (2022) 137437}
  [\href{https://arxiv.org/abs/2203.01856}{{\ttfamily 2203.01856}}].

\bibitem{Wu:2021ndf}
Q.~Wu, T.~Zhu, R.~Niu, W.~Zhao and A.~Wang, \emph{{Constraints on the Nieh-Yan
  modified teleparallel gravity with gravitational waves}},
  \href{https://doi.org/10.1103/PhysRevD.105.024035}{\emph{Phys. Rev. D}
  {\bfseries 105} (2022) 024035}
  [\href{https://arxiv.org/abs/2110.13870}{{\ttfamily 2110.13870}}].

\bibitem{Martin-Ruiz:2017cjt}
A.~Mart\'\i{}n-Ruiz and L.F.~Urrutia, \emph{{Gravitational waves propagation in
  nondynamical Chern\textendash{}Simons gravity}},
  \href{https://doi.org/10.1142/S0218271817501486}{\emph{Int. J. Mod. Phys. D}
  {\bfseries 26} (2017) 1750148}
  [\href{https://arxiv.org/abs/1706.08843}{{\ttfamily 1706.08843}}].

\bibitem{Nojiri:2019nar}
S.~Nojiri, S.D.~Odintsov, V.K.~Oikonomou and A.A.~Popov, \emph{{Propagation of
  Gravitational Waves in Chern-Simons Axion Einstein Gravity}},
  \href{https://doi.org/10.1103/PhysRevD.100.084009}{\emph{Phys. Rev. D}
  {\bfseries 100} (2019) 084009}
  [\href{https://arxiv.org/abs/1909.01324}{{\ttfamily 1909.01324}}].

\bibitem{Nojiri:2020pqr}
S.~Nojiri, S.D.~Odintsov, V.K.~Oikonomou and A.A.~Popov, \emph{{Propagation of
  gravitational waves in Chern\textendash{}Simons axion $F(R)$ gravity}},
  \href{https://doi.org/10.1016/j.dark.2020.100514}{\emph{Phys. Dark Univ.}
  {\bfseries 28} (2020) 100514}
  [\href{https://arxiv.org/abs/2002.10402}{{\ttfamily 2002.10402}}].

\bibitem{Boudet:2022nub}
S.~Boudet, F.~Bombacigno, F.~Moretti and G.J.~Olmo, \emph{{Torsional
  birefringence in metric-affine Chern-Simons gravity: gravitational waves in
  late-time cosmology}},  \href{https://arxiv.org/abs/2209.14394}{{\ttfamily
  2209.14394}}.

\bibitem{Li:2022grj}
Z.~Li, J.~Qiao, T.~Liu, T.~Zhu and W.~Zhao, \emph{{Gravitational Waveform and
  Polarization from Binary Black Hole Inspiral in Dynamical Chern-Simons
  Gravity: From Generation to Propagation}},
  \href{https://arxiv.org/abs/2211.12188}{{\ttfamily 2211.12188}}.

\bibitem{ALEXANDER2008444}
S.H.~Alexander, \emph{Is cosmic parity violation responsible for the anomalies
  in the wmap data?},
  \href{https://doi.org/https://doi.org/10.1016/j.physletb.2007.12.010}{\emph{Physics
  Letters B} {\bfseries 660} (2008) 444}.

\bibitem{PhysRevLett.83.1506}
A.~Lue, L.~Wang and M.~Kamionkowski, \emph{Cosmological signature of new
  parity-violating interactions},
  \href{https://doi.org/10.1103/PhysRevLett.83.1506}{\emph{Phys. Rev. Lett.}
  {\bfseries 83} (1999) 1506}.

\bibitem{Bartolo:2018elp}
N.~Bartolo, G.~Orlando and M.~Shiraishi, \emph{{Measuring chiral gravitational
  waves in Chern-Simons gravity with CMB bispectra}},
  \href{https://doi.org/10.1088/1475-7516/2019/01/050}{\emph{JCAP} {\bfseries
  01} (2019) 050} [\href{https://arxiv.org/abs/1809.11170}{{\ttfamily
  1809.11170}}].

\bibitem{Bartolo:2017szm}
N.~Bartolo and G.~Orlando, \emph{{Parity breaking signatures from a
  Chern-Simons coupling during inflation: the case of non-Gaussian
  gravitational waves}},
  \href{https://doi.org/10.1088/1475-7516/2017/07/034}{\emph{JCAP} {\bfseries
  07} (2017) 034} [\href{https://arxiv.org/abs/1706.04627}{{\ttfamily
  1706.04627}}].

\bibitem{PhysRevD.99.064049}
S.D.~Odintsov and V.K.~Oikonomou, \emph{$f(r)$ gravity inflation with
  string-corrected axion dark matter},
  \href{https://doi.org/10.1103/PhysRevD.99.064049}{\emph{Phys. Rev. D}
  {\bfseries 99} (2019) 064049}.

\bibitem{Odintsov:2021kup}
S.D.~Odintsov, V.K.~Oikonomou and F.P.~Fronimos, \emph{{Quantitative
  predictions for f(R) gravity primordial gravitational waves}},
  \href{https://doi.org/10.1016/j.dark.2022.100950}{\emph{Phys. Dark Univ.}
  {\bfseries 35} (2022) 100950}
  [\href{https://arxiv.org/abs/2108.11231}{{\ttfamily 2108.11231}}].

\bibitem{sym14040729}
S.D.~Odintsov, V.K.~Oikonomou and R.~Myrzakulov, \emph{Spectrum of primordial
  gravitational waves in modified gravities: A short overview},
  \href{https://doi.org/10.3390/sym14040729}{\emph{Symmetry} {\bfseries 14}
  (2022) }.

\bibitem{PhysRevD.105.104054}
S.D.~Odintsov and V.K.~Oikonomou, \emph{Chirality of gravitational waves in
  chern-simons $f(r)$ gravity cosmology},
  \href{https://doi.org/10.1103/PhysRevD.105.104054}{\emph{Phys. Rev. D}
  {\bfseries 105} (2022) 104054}.

\bibitem{Satoh:2007gn}
M.~Satoh, S.~Kanno and J.~Soda, \emph{{Circular Polarization of Primordial
  Gravitational Waves in String-inspired Inflationary Cosmology}},
  \href{https://doi.org/10.1103/PhysRevD.77.023526}{\emph{Phys. Rev. D}
  {\bfseries 77} (2008) 023526}
  [\href{https://arxiv.org/abs/0706.3585}{{\ttfamily 0706.3585}}].

\bibitem{Takahashi:2009wc}
T.~Takahashi and J.~Soda, \emph{{Chiral Primordial Gravitational Waves from a
  Lifshitz Point}},
  \href{https://doi.org/10.1103/PhysRevLett.102.231301}{\emph{Phys. Rev. Lett.}
  {\bfseries 102} (2009) 231301}
  [\href{https://arxiv.org/abs/0904.0554}{{\ttfamily 0904.0554}}].

\bibitem{Kamada:2021kxi}
K.~Kamada, J.~Kume and Y.~Yamada, \emph{{Chiral gravitational effect in
  time-dependent backgrounds}},
  \href{https://doi.org/10.1007/JHEP05(2021)292}{\emph{JHEP} {\bfseries 05}
  (2021) 292} [\href{https://arxiv.org/abs/2104.00583}{{\ttfamily
  2104.00583}}].

\bibitem{PhysRevLett.96.081301}
S.H.S.~Alexander, M.E.~Peskin and M.M.~Sheikh-Jabbari, \emph{Leptogenesis from
  gravity waves in models of inflation},
  \href{https://doi.org/10.1103/PhysRevLett.96.081301}{\emph{Phys. Rev. Lett.}
  {\bfseries 96} (2006) 081301}.

\bibitem{PhysRevD.69.023504}
J.~Garc\'{\i}a-Bellido, M.~Garc\'{\i}a~P\'erez and A.~Gonz\'alez-Arroyo,
  \emph{Chern-simons production during preheating in hybrid inflation models},
  \href{https://doi.org/10.1103/PhysRevD.69.023504}{\emph{Phys. Rev. D}
  {\bfseries 69} (2004) 023504}.

\bibitem{Alexander_2006}
S.H.S.~Alexander and S.J.~Gates, \emph{Can the string scale be related to the
  cosmic baryon asymmetry?},
  \href{https://doi.org/10.1088/1475-7516/2006/06/018}{\emph{Journal of
  Cosmology and Astroparticle Physics} {\bfseries 2006} (2006) 018}.

\bibitem{Jimenez:2017cdr}
D.~Jim\'enez, K.~Kamada, K.~Schmitz and X.-J.~Xu, \emph{{Baryon asymmetry and
  gravitational waves from pseudoscalar inflation}},
  \href{https://doi.org/10.1088/1475-7516/2017/12/011}{\emph{JCAP} {\bfseries
  12} (2017) 011} [\href{https://arxiv.org/abs/1707.07943}{{\ttfamily
  1707.07943}}].

\bibitem{PhysRevD.80.064008}
V.~Cardoso and L.~Gualtieri, \emph{Perturbations of schwarzschild black holes
  in dynamical chern-simons modified gravity},
  \href{https://doi.org/10.1103/PhysRevD.80.064008}{\emph{Phys. Rev. D}
  {\bfseries 80} (2009) 064008}.

\bibitem{PhysRevD.81.089903}
V.~Cardoso and L.~Gualtieri, \emph{Erratum: Perturbations of schwarzschild
  black holes in dynamical chern-simons modified gravity [phys. rev. d 80,
  064008 (2009)]},
  \href{https://doi.org/10.1103/PhysRevD.81.089903}{\emph{Phys. Rev. D}
  {\bfseries 81} (2010) 089903}.

\bibitem{PhysRevD.81.124021}
C.~Molina, P.~Pani, V.~Cardoso and L.~Gualtieri, \emph{Gravitational signature
  of schwarzschild black holes in dynamical chern-simons gravity},
  \href{https://doi.org/10.1103/PhysRevD.81.124021}{\emph{Phys. Rev. D}
  {\bfseries 81} (2010) 124021}.

\bibitem{PhysRevD.80.064006}
C.F.~Sopuerta and N.~Yunes, \emph{Extreme- and intermediate-mass ratio
  inspirals in dynamical chern-simons modified gravity},
  \href{https://doi.org/10.1103/PhysRevD.80.064006}{\emph{Phys. Rev. D}
  {\bfseries 80} (2009) 064006}.

\bibitem{Yunes:2010yf}
N.~Yunes, R.~O'Shaughnessy, B.J.~Owen and S.~Alexander, \emph{{Testing
  gravitational parity violation with coincident gravitational waves and short
  gamma-ray bursts}},
  \href{https://doi.org/10.1103/PhysRevD.82.064017}{\emph{Phys. Rev. D}
  {\bfseries 82} (2010) 064017}
  [\href{https://arxiv.org/abs/1005.3310}{{\ttfamily 1005.3310}}].

\bibitem{Cano:2019ore}
P.A.~Cano and A.~Ruip\'erez, \emph{{Leading higher-derivative corrections to
  Kerr geometry}}, \href{https://doi.org/10.1007/JHEP05(2019)189}{\emph{JHEP}
  {\bfseries 05} (2019) 189}
  [\href{https://arxiv.org/abs/1901.01315}{{\ttfamily 1901.01315}}].

\bibitem{Harko:2009kj}
T.~Harko, Z.~Kovacs and F.S.N.~Lobo, \emph{{Thin accretion disk signatures in
  dynamical Chern-Simons modified gravity}},
  \href{https://doi.org/10.1088/0264-9381/27/10/105010}{\emph{Class. Quant.
  Grav.} {\bfseries 27} (2010) 105010}
  [\href{https://arxiv.org/abs/0909.1267}{{\ttfamily 0909.1267}}].

\bibitem{Ghodrati:2017roz}
H.~Motohashi and T.~Suyama, \emph{{Black hole perturbation in parity violating
  gravitational theories}},
  \href{https://doi.org/10.1103/PhysRevD.84.084041}{\emph{Phys. Rev. D}
  {\bfseries 84} (2011) 084041}
  [\href{https://arxiv.org/abs/1107.3705}{{\ttfamily 1107.3705}}].

\bibitem{Motohashi:2011pw}
H.~Motohashi and T.~Suyama, \emph{{Black hole perturbation in parity violating
  gravitational theories}},
  \href{https://doi.org/10.1103/PhysRevD.84.084041}{\emph{Phys. Rev. D}
  {\bfseries 84} (2011) 084041}
  [\href{https://arxiv.org/abs/1107.3705}{{\ttfamily 1107.3705}}].

\bibitem{Yagi:2017zhb}
K.~Yagi and H.~Yang, \emph{{Probing Gravitational Parity Violation with
  Gravitational Waves from Stellar-mass Black Hole Binaries}},
  \href{https://doi.org/10.1103/PhysRevD.97.104018}{\emph{Phys. Rev. D}
  {\bfseries 97} (2018) 104018}
  [\href{https://arxiv.org/abs/1712.00682}{{\ttfamily 1712.00682}}].

\bibitem{Wagle:2021tam}
P.~Wagle, N.~Yunes and H.O.~Silva, \emph{{Quasinormal modes of slowly-rotating
  black holes in dynamical Chern-Simons gravity}},
  \href{https://doi.org/10.1103/PhysRevD.105.124003}{\emph{Phys. Rev. D}
  {\bfseries 105} (2022) 124003}
  [\href{https://arxiv.org/abs/2103.09913}{{\ttfamily 2103.09913}}].

\bibitem{Iosifidis:2020dck}
D.~Iosifidis and L.~Ravera, \emph{{Parity Violating Metric-Affine Gravity
  Theories}},  \href{https://arxiv.org/abs/2009.03328}{{\ttfamily 2009.03328}}.

\bibitem{Li:2022vtn}
M.~Li, Y.~Tong and D.~Zhao, \emph{{Possible consistent model of parity
  violations in the symmetric teleparallel gravity}},
  \href{https://doi.org/10.1103/PhysRevD.105.104002}{\emph{Phys. Rev. D}
  {\bfseries 105} (2022) 104002}
  [\href{https://arxiv.org/abs/2203.06912}{{\ttfamily 2203.06912}}].

\bibitem{Jackiw:2003pm}
R.~Jackiw and S.Y.~Pi, \emph{{Chern-Simons modification of general
  relativity}}, \href{https://doi.org/10.1103/PhysRevD.68.104012}{\emph{Phys.
  Rev. D} {\bfseries 68} (2003) 104012}
  [\href{https://arxiv.org/abs/gr-qc/0308071}{{\ttfamily gr-qc/0308071}}].

\bibitem{Alexander:2009tp}
S.~Alexander and N.~Yunes, \emph{{Chern-Simons Modified General Relativity}},
  \href{https://doi.org/10.1016/j.physrep.2009.07.002}{\emph{Phys. Rept.}
  {\bfseries 480} (2009) 1} [\href{https://arxiv.org/abs/0907.2562}{{\ttfamily
  0907.2562}}].

\bibitem{Boudet:2022wmb}
S.~Boudet, F.~Bombacigno, G.J.~Olmo and P.J.~Porfirio, \emph{{Quasinormal modes
  of Schwarzschild black holes in projective invariant Chern-Simons modified
  gravity}}, \href{https://doi.org/10.1088/1475-7516/2022/05/032}{\emph{JCAP}
  {\bfseries 05} (2022) 032}
  [\href{https://arxiv.org/abs/2203.04000}{{\ttfamily 2203.04000}}].

\bibitem{Sulantay:2022sag}
F.~Sulantay, M.~Lagos and M.~Ba\~nados, \emph{{Chiral Gravitational Waves in
  Palatini Chern-Simons}},  \href{https://arxiv.org/abs/2211.08925}{{\ttfamily
  2211.08925}}.

\bibitem{Kostelecky:2003fs}
V.A.~Kostelecky, \emph{{Gravity, Lorentz violation, and the standard model}},
  \href{https://doi.org/10.1103/PhysRevD.69.105009}{\emph{Phys. Rev. D}
  {\bfseries 69} (2004) 105009}
  [\href{https://arxiv.org/abs/hep-th/0312310}{{\ttfamily hep-th/0312310}}].

\bibitem{Bluhm:2004ep}
R.~Bluhm and V.A.~Kostelecky, \emph{{Spontaneous Lorentz violation,
  Nambu-Goldstone modes, and gravity}},
  \href{https://doi.org/10.1103/PhysRevD.71.065008}{\emph{Phys. Rev. D}
  {\bfseries 71} (2005) 065008}
  [\href{https://arxiv.org/abs/hep-th/0412320}{{\ttfamily hep-th/0412320}}].

\bibitem{Bailey:2006fd}
Q.G.~Bailey and V.A.~Kostelecky, \emph{{Signals for Lorentz violation in
  post-Newtonian gravity}},
  \href{https://doi.org/10.1103/PhysRevD.74.045001}{\emph{Phys. Rev. D}
  {\bfseries 74} (2006) 045001}
  [\href{https://arxiv.org/abs/gr-qc/0603030}{{\ttfamily gr-qc/0603030}}].

\bibitem{Bluhm:2007bd}
R.~Bluhm, S.-H.~Fung and V.A.~Kostelecky, \emph{{Spontaneous Lorentz and
  Diffeomorphism Violation, Massive Modes, and Gravity}},
  \href{https://doi.org/10.1103/PhysRevD.77.065020}{\emph{Phys. Rev. D}
  {\bfseries 77} (2008) 065020}
  [\href{https://arxiv.org/abs/0712.4119}{{\ttfamily 0712.4119}}].

\bibitem{Delhom:2019wcm}
A.~Delhom, J.R.~Nascimento, G.J.~Olmo, A.Y.~Petrov and P.J.~Porf\'\i{}rio,
  \emph{{Metric-affine bumblebee gravity: classical aspects}},
  \href{https://doi.org/10.1140/epjc/s10052-021-09083-y}{\emph{Eur. Phys. J. C}
  {\bfseries 81} (2021) 287}
  [\href{https://arxiv.org/abs/1911.11605}{{\ttfamily 1911.11605}}].

\bibitem{Delhom:2020gfv}
A.~Delhom, J.R.~Nascimento, G.J.~Olmo, A.Y.~Petrov and P.J.~Porf\'\i{}rio,
  \emph{{Radiative corrections in metric-affine bumblebee model}},
  \href{https://doi.org/10.1016/j.physletb.2022.136932}{\emph{Phys. Lett. B}
  {\bfseries 826} (2022) 136932}
  [\href{https://arxiv.org/abs/2010.06391}{{\ttfamily 2010.06391}}].

\bibitem{Delhom:2022xfo}
A.~Delhom, T.~Mariz, J.R.~Nascimento, G.J.~Olmo, A.Y.~Petrov and P.J.~Porfirio,
  \emph{{Spontaneous Lorentz symmetry breaking and one-loop effective action in
  the metric-affine bumblebee gravity}},
  \href{https://arxiv.org/abs/2202.11613}{{\ttfamily 2202.11613}}.

\bibitem{Zhu:2013fja}
T.~Zhu, W.~Zhao, Y.~Huang, A.~Wang and Q.~Wu, \emph{{Effects of parity
  violation on non-gaussianity of primordial gravitational waves in
  Ho\v{r}ava-Lifshitz gravity}},
  \href{https://doi.org/10.1103/PhysRevD.88.063508}{\emph{Phys. Rev. D}
  {\bfseries 88} (2013) 063508}
  [\href{https://arxiv.org/abs/1305.0600}{{\ttfamily 1305.0600}}].

\bibitem{Gong:2021jgg}
C.~Gong, T.~Zhu, R.~Niu, Q.~Wu, J.-L.~Cui, X.~Zhang et~al.,
  \emph{{Gravitational wave constraints on Lorentz and parity violations in
  gravity: High-order spatial derivative cases}},
  \href{https://doi.org/10.1103/PhysRevD.105.044034}{\emph{Phys. Rev. D}
  {\bfseries 105} (2022) 044034}
  [\href{https://arxiv.org/abs/2112.06446}{{\ttfamily 2112.06446}}].

\bibitem{Li:2020xjt}
M.~Li, H.~Rao and D.~Zhao, \emph{{A simple parity violating gravity model
  without ghost instability}},
  \href{https://doi.org/10.1088/1475-7516/2020/11/023}{\emph{JCAP} {\bfseries
  11} (2020) 023} [\href{https://arxiv.org/abs/2007.08038}{{\ttfamily
  2007.08038}}].

\bibitem{Li:2021wij}
M.~Li, H.~Rao and Y.~Tong, \emph{{Revisiting a parity violating gravity model
  without ghost instability: Local Lorentz covariance}},
  \href{https://doi.org/10.1103/PhysRevD.104.084077}{\emph{Phys. Rev. D}
  {\bfseries 104} (2021) 084077}
  [\href{https://arxiv.org/abs/2104.05917}{{\ttfamily 2104.05917}}].

\bibitem{Moretti2019}
F.~Moretti, F.~Bombacigno and G.~Montani, \emph{{Gauge invariant formulation of
  metric f (R) gravity for gravitational waves}},
  \href{https://doi.org/10.1103/PhysRevD.100.084014}{\emph{Physical Review D}
  {\bfseries 100} (2019) 084014}.

\bibitem{Landau:1946jc}
L.D.~Landau, \emph{{On the vibrations of the electronic plasma}}, {\emph{J.
  Phys. (USSR)} {\bfseries 10} (1946) 25}.

\bibitem{lifshitz1995physical}
E.~Lifshitz and L.~Pitaevskii, \emph{Physical Kinetics: Volume 10}, Course of
  theoretical physics, Elsevier Science (1995).

\bibitem{Montani:2018iqd}
G.~Montani and F.~Moretti, \emph{{Modified Gravitational Waves Across Galaxies
  from Macroscopic Gravity}},
  \href{https://doi.org/10.1103/PhysRevD.100.024045}{\emph{Phys. Rev. D}
  {\bfseries 100} (2019) 024045}
  [\href{https://arxiv.org/abs/1805.08018}{{\ttfamily 1805.08018}}].

\bibitem{stix1992waves}
T.~Stix, \emph{Waves in Plasmas}, American Inst. of Physics (1992).

\bibitem{Hehl:1994ue}
F.W.~Hehl, J.D.~McCrea, E.W.~Mielke and Y.~Ne'eman, \emph{{Metric affine gauge
  theory of gravity: Field equations, Noether identities, world spinors, and
  breaking of dilation invariance}},
  \href{https://doi.org/10.1016/0370-1573(94)00111-F}{\emph{Phys. Rept.}
  {\bfseries 258} (1995) 1}
  [\href{https://arxiv.org/abs/gr-qc/9402012}{{\ttfamily gr-qc/9402012}}].

\bibitem{Iosifidis2021}
D.~Iosifidis and E.N.~Saridakis, \emph{Metric-affine gravity},  in
  \emph{Modified Gravity and Cosmology: An Update by the CANTATA Network},
  E.N.~Saridakis, R.~Lazkoz, V.~Salzano, P.V.~Moniz, S.~Capozziello,
  J.~Beltr{\'a}n~Jim{\'e}nez et~al., eds., (Cham), pp.~129--142, Springer
  International Publishing (2021),
  \href{https://doi.org/10.1007/978-3-030-83715-0_10}{DOI}.

\bibitem{Jimenez:2022hcz}
J.B.~Jim\'enez, A.~Jim\'enez-Cano and Y.N.~Obukhov, \emph{{On parity-odd sector
  in metric-affine theories}},
  \href{https://arxiv.org/abs/2210.01729}{{\ttfamily 2210.01729}}.

\bibitem{sofue2020rotation}
Y.~Sofue, \emph{Rotation curve of the milky way and the dark matter density},
  {\emph{Galaxies} {\bfseries 8} (2020) 37}.

\bibitem{2020PhRvD.102f2003B}
A.~{Buikema}, C.~{Cahillane}, G.L.~{Mansell}, C.D.~{Blair}, R.~{Abbott},
  C.~{Adams} et~al., \emph{{Sensitivity and performance of the Advanced LIGO
  detectors in the third observing run}},
  \href{https://doi.org/10.1103/PhysRevD.102.062003}{\emph{Phys. Rev. D}
  {\bfseries 102} (2020) 062003}
  [\href{https://arxiv.org/abs/2008.01301}{{\ttfamily 2008.01301}}].

\bibitem{2021Univ....7..322B}
D.~{Bersanetti}, B.~{Patricelli}, O.J.~{Piccinni}, F.~{Piergiovanni},
  F.~{Salemi} and V.~{Sequino}, \emph{{Advanced Virgo: Status of the Detector,
  Latest Results and Future Prospects}},
  \href{https://doi.org/10.3390/universe7090322}{\emph{Universe} {\bfseries 7}
  (2021) 322}.

\bibitem{2019CQGra..36j5011R}
T.~{Robson}, N.J.~{Cornish} and C.~{Liu}, \emph{{The construction and use of
  LISA sensitivity curves}},
  \href{https://doi.org/10.1088/1361-6382/ab1101}{\emph{Classical and Quantum
  Gravity} {\bfseries 36} (2019) 105011}
  [\href{https://arxiv.org/abs/1803.01944}{{\ttfamily 1803.01944}}].

\bibitem{Nakamura:2018yaw}
Y.~Nakamura, D.~Kikuchi, K.~Yamada, H.~Asada and N.~Yunes,
  \emph{{Weakly-gravitating objects in dynamical Chern\textendash{}Simons
  gravity and constraints with gravity probe B}},
  \href{https://doi.org/10.1088/1361-6382/ab04c5}{\emph{Class. Quant. Grav.}
  {\bfseries 36} (2019) 105006}
  [\href{https://arxiv.org/abs/1810.13313}{{\ttfamily 1810.13313}}].

\bibitem{Guarrera:2007tu}
D.~Guarrera and A.J.~Hariton, \emph{{Papapetrou Energy-Momentum Tensor for
  Chern-Simons Modified Gravity}},
  \href{https://doi.org/10.1103/PhysRevD.76.044011}{\emph{Phys. Rev. D}
  {\bfseries 76} (2007) 044011}
  [\href{https://arxiv.org/abs/gr-qc/0702029}{{\ttfamily gr-qc/0702029}}].

\bibitem{Stein:2010pn}
L.C.~Stein and N.~Yunes, \emph{{Effective Gravitational Wave Stress-energy
  Tensor in Alternative Theories of Gravity}},
  \href{https://doi.org/10.1103/PhysRevD.83.064038}{\emph{Phys. Rev. D}
  {\bfseries 83} (2011) 064038}
  [\href{https://arxiv.org/abs/1012.3144}{{\ttfamily 1012.3144}}].

\bibitem{Bhattacharyya:2018hsj}
S.~Bhattacharyya and S.~Shankaranarayanan, \emph{{Distinguishing general
  relativity from Chern-Simons gravity using gravitational wave
  polarizations}},
  \href{https://doi.org/10.1103/PhysRevD.100.024022}{\emph{Phys. Rev. D}
  {\bfseries 100} (2019) 024022}
  [\href{https://arxiv.org/abs/1812.00187}{{\ttfamily 1812.00187}}].

\end{thebibliography}\endgroup
\end{document}